\documentclass[oldversion]{aa}  

\usepackage{graphicx}
\usepackage{txfonts}
\usepackage{natbib}

\def\ffs{\hbox{$\,.\!\!^{\rm s}$}}
\def\ffas{\hbox{$\,.\!\!^{\prime\prime}$}}

\begin{document}
%



\title{High-resolution mapping of the physical conditions in two nearby active galaxies based on $^{12}$CO(1--0), (2--1) and (3--2) lines	\thanks{Based on observations carried out with the Submillimeter Array (SMA), a joint project between the Smithsonian Astrophysical Observatory and the Academia Sinica Institute of Astronomy and Astrophysics that is funded by the Smithsonian Institution and the Academia Sinica.}
	\thanks{Based on observations carried out with the IRAM Plateau de Bure
 Interferometer. IRAM is supported by INSU/CNRS (France), MPG (Germany) and IGN (Spain).}
 }

  \titlerunning{Mapping the physical conditions of the molecular gas in the central regions of two active galaxies}

   \author{F.\,Boone
          \inst{1}$^{,}$\inst{2}
       \and
       S.\,Garc\'\i a-Burillo\inst{3}
	   \and
	   F.\,Combes\inst{1}
	   \and
	   J.\,Lim\inst{4}
	   \and 
	   P. Ho\inst{4}
	   \and 
	   A.\,J.\,Baker\inst{5}
	   \and
	   S. Matsushita\inst{4}$^{,}$\inst{6}
	   \and
	   M. Krips\inst{7}
	   \and
	   Dinh-van-Trung\inst{4}
	   \and
	   E. Schinnerer\inst{8}	   
          }

   \offprints{F.\,Boone}

   \institute{Observatoire de Paris, LERMA, 
              61 avenue de l'Observatoire, F-75014 Paris, France\\
              \email{frederic.boone@obspm.fr}
         \and
              Laboratoire d'Astrophysique de Toulouse-Tarbes, Universit\'e de Toulouse, CNRS, 14 Avenue Edouard Belin, F-31400 Toulouse, France
         \and
              Observatorio Astron\'omico Nacional (OAN) - Observatorio de Madrid, C/ Alfonso XII 3, 28014 Madrid, Spain
         \and
         Institute of Astronomy and Astrophysics, Academia Sinica, P.O. Box 23-141, Taipei 106, Taiwan
	      \and
        Department of Physics and Astronomy, Rutgers, the State University of 
New Jersey, 136 Frelinghuysen Road, Piscataway, NJ 08854, USA
	      \and
	        Joint ALMA Office, Av. El Golf 40, Piso 18, Las Condes, Santiago, Chile 
	      \and
        Institut de Radio Astronomie Millim\'etrique, 
        300 Rue de la Piscine, 38406 Saint Martin d'H\`eres, France
	\and
	Max-Planck-Institut f\"ur Astronomie, 
	K\"onigstuhl 17, 69117 Heidelberg, Germany
             }

   \date{accepted 7 October 2010}

 
  \abstract{We present a detailed analysis of high resolution observations of the three lowest CO transitions in two nearby active galaxies, NGC\,4569 and NGC\,4826.  The CO(1--0) and (2--1) lines were observed with the Plateau de Bure Interferometer and the CO(3--2) line with the Submillimeter Array.
Combining these data allows us to compare the emission in the three lines and to map the line ratios, $R_{21}=I_{\rm CO(2-1)}/I_{\rm CO(1-0)}$ and $R_{32}=I_{\rm CO(3-2)}/I_{\rm CO(1-0)}$ at a resolution of $\sim$2$''$, i.e. a linear resolution of 160\,pc for NGC\,4569 and 40\,pc for NGC\,4826. 
In both galaxies the emission in the three lines is similarly distributed spatially and in velocity, and CO is less excited ($R_{32}<0.6)$ than in the Galactic Center or the centers of other active galaxies studied so far. 
According to a pseudo-LTE model the molecular gas in NGC\,4569 is cold and mainly optically thick in the CO(1--0) and (2--1)  lines; less than 50\% of the gas is optically thin in the CO(3--2) line. LVG modeling suggests the presence of an elongated ring of cold and dense gas coinciding with the inner Lindblad resonance (ILR) of the stellar bar in agreement with a previous analysis of the kinematics.  More excited gas is resolved in the circumnuclear disk of NGC\,4826. According to our pseudo-LTE model this corresponds to warmer gas with a $\sim$50$\%$ of the CO(3--2) emission being optically thin. LVG modeling indicates the presence of a semicircular arc of dense and cold gas centered on the dynamical center and $\sim$70\,pc in radius. The gas temperature increases and its density decreases toward the center.
 A near side/far side asymmetry noticeable in the CO, $R_{32}$ and Pa$\alpha$ maps suggests that opacity effects play a role. Examining published CO maps of nearby active galaxies we find similar asymmetries suggesting that this could be a common phenomenon in active galaxies.  These mainly qualitative results
 open new perspectives for the study of active galaxies with the future Atacama Large Millimeter/submillimeter Array. }
 
 \keywords{galaxies: individual: NGC\,4569, NGC\,4826 -- galaxies: active}

   \maketitle
%
\defcitealias{2007A&A...471..113B}{NUGA-VII}
\defcitealias{2003A&A...407..485G}{NUGA-I}
\defcitealias{2005A&A...441.1011G}{NUGA-IV}
\section{Introduction}
Multitransition CO observations are essential for studying the processes at work in the centers of active galaxies. The molecular gas fuels the  nuclear activity powered by accretion onto a supermassive black hole and/or intense star formation. Conversely, the molecular gas properties are affected by the influence of the active nucleus through jets, winds, and radiation. The molecular gas must also constitute the bulk of the nuclear torus responsible for the highly anisotropic extinction invoked by the unified model for active galactic nuclei \citep[AGN; ][]{1993ARA&A..31..473A}. As the most abundant molecule after H$_2$ and with its four lowest rotational transitions coinciding with the atmospheric windows in the millimeter and submillimeter range, CO has long been recognized and used as the best molecular gas tracer. 
Observing a single transition allows us to map the molecular gas column density and kinematics and, thus, to identify dynamical instabilities, and measure gravitational torques and possible gas inflows. Such detailed studies are for example conducted for nearby galaxies in the context of the NUclei of GAlaxies survey \citep[][ hereafter NUGA-I]{2003A&A...407..485G} based on high resolution observations of the CO(1--0) and (2--1) lines \citep[see also][for other high resolution CO surveys]{2003ApJS..145..259H,1999ApJS..124..403S}.  It is however necessary to observe more transitions to characterize the physical conditions of the molecular gas and identify interstellar medium (ISM) structures involved in star formation, shocks (e.g., due to the disk kinematics, AGN feedback, stellar winds, or supernova remnants), or quiescent gas reservoirs ready to ignite star formation or feed the central black hole. 
Other molecular  or atomic lines { (including those of CO isotopologues)} can be observed to characterize the physical conditions, but this implies adding new unknowns, namely the relative abundances of the observed molecules or atoms (which are otherwise interesting to measure to study the ISM chemistry). 

Models of the ISM in galactic centers suggest that high resolution multitransition CO observations should allow for the identification of the  torus around the supermassive black hole \citep{2005ApJ...619...93W} as well as the influence of the AGN on the ISM chemistry  \citep[in X-ray dominated regions, ][]{2007A&A...461..793M,2005A&A...436..397M}. 

While the first two CO transitions at 115 and 230\,GHz have been regularly observed for some time at high spatial resolution with millimeter interferometers, the third transition at 345\,GHz could only be observed with single dish telescopes until recently. CO(3--2) single 
dish observations complemented  lower transitions' observations and yielded a wealth of information on the excitation of the molecular gas in galaxy centers \citep[see e.g. ][]{1999ApJ...516..114P,2000ApJ...537..631P,2001A&A...373..853D,2001A&A...367..457C,2003A&A...398..959H, 2003ApJ...588..771Y,2005A&A...438..533W,2006A&A...460..467B, 2007PASJ...59...43M, 2008PASJ...60..457K,2008A&A...491..483P, 2009ApJ...693.1736W,2009A&A...493..525I} including that of our own Galaxy \citep{2007PASJ...59...15O}. These observations showed that the CO(3--2)/(1--0) line ratio generally increases toward the center, where it can reach values greater than unity (in temperature units) in starburst and AGN hosts as well as in the Milky Way (vs.\ a typical value of 0.4 in galaxy disks). It is only since the Submillimeter Array became operational \citep{2004ApJ...616L...1H} that it has been possible to map the CO(3--2)  \citep[and higher $J$-transitions see, e.g., ][]{2009ApJ...693...56M} in the central regions of galaxies with the same spatial resolution as the CO(1--0) and CO(2--1) lines observed with millimeter interferometers. Among other results, these observations have revealed gas interacting with a jet in the center of M\,51 \citep{2004ApJ...616L..55M} and gas interacting with a starburst superwind or AGN outflow in NGC\,6240  \citep{2007ApJ...659..283I}. 

In this paper, we present a detailed analysis of high resolution observations of the three lowest CO transitions in two nearby active galaxies. Long-term drivers for this work are (1) to estimate the feasibility of mapping the physical conditions in the centers of galaxies based on CO line observations, and (2) to discuss the implications for the study of active galaxies with current and future instruments. We selected the two galaxies in the NUGA sample with the strongest low-$J$ CO lines, NGC\,4569 and NGC\,4826, and observed the CO(3--2) line with the { Submillimeter Array (SMA)}. The CO(1--0) and CO(2--1) observations obtained with the IRAM Plateau de Bure Interferometer (PdBI) were presented and analyzed in  \citetalias{2003A&A...407..485G}, \citet[][ hereafter NUGA-IV]{2005A&A...441.1011G} and \citet[][ hereafter NUGA-VII]{2007A&A...471..113B}.

NGC\,4569 is a bright early type spiral galaxy in the Virgo cluster at a distance of 17 Mpc (1$''$ is $\sim$80\,pc). It harbours a  transition-type nucleus \citep{1997ApJS..112..315H} which exhibits the most pronounced nuclear starburst activity among the { low-ionization nuclear emission-line region (LINER)} and transition nuclei with available UV data \citep{1998AJ....116...55M}.  \citet{2003IAUJD..10E..38H} report a bipolar outflow in X-ray and a giant H$\alpha$ outflow with a scale height of 10 kpc to the west. \citet{2006A&A...447..465C} observe symmetric radio lobes extending up to 24 kpc from the galactic disk. High resolution PdBI observations allowed us to model the kinematics of the central kpc \citepalias{2007A&A...471..113B} and showed that the molecular gas concentration at a radius of 600 pc corresponds to the inner Lindblad resonance (ILR) of the large scale bar, which extends to a radius of 5 kpc \citep{2002MNRAS.337.1118L}.
The inclination of the galaxy is 70\,$\deg$ \citep{1988ngc..book.....T} and its position angle 30\,$\deg$ \citep{1988AJ.....96..851G}.
%

NGC\,4826, also known as the 'Evil Eye', is the closest target (4.1 Mpc, i.e. 1$''$ is $\sim$20\,pc) of the  NUGA core sample for which we have acquired $\sim$0.5'' resolution PdBI CO observations \citepalias{2005A&A...441.1011G}. These observations show a large concentration of molecular gas within a radius of 80\,pc forming a circumnuclear molecular disk (CND). A detailed analysis of the kinematics, however, does not reveal any evidence of fuelling of the nucleus \citepalias{2003A&A...407..485G}. The nucleus of NGC\,4826 is also classified as a transition type \citep{1997ApJS..112..315H}. The inclination of the galaxy is 60\,$\deg$ \citep{1994AJ....107..173R}, and its position angle 112\,$\deg$ \citepalias[][]{2003A&A...407..485G}.


The paper is organized as follows. The SMA CO(3--2) observations are presented in Section\,\ref{sec:observations}. The distributions of the emission in the three lines and of the two line ratios { ($R_{21}\equiv I_{\rm CO(2-1)}/I_{\rm CO(1-0)}$ and $R_{32}\equiv I_{\rm CO(3-2)}/I_{\rm CO(1-0)}$, where $I_{\rm CO}$ is a line intensity in temperature units)} are compared in Section\,\ref{sec:3lines} and Section\,\ref{sec:lineratios}, respectively.  Mapping of the physical conditions based on radiative transfer modeling of the line ratios is performed in Section\,\ref{sec:physcond}.  The results obtained are discussed in Section\,\ref{sec:discussion} and conclusions are drawn in Section\,\ref{sec:conclusion}.

\section{Observations and data processing}\label{sec:observations}

\begin{figure*}
\centering
\resizebox{16cm}{!}{\includegraphics[width=7cm]{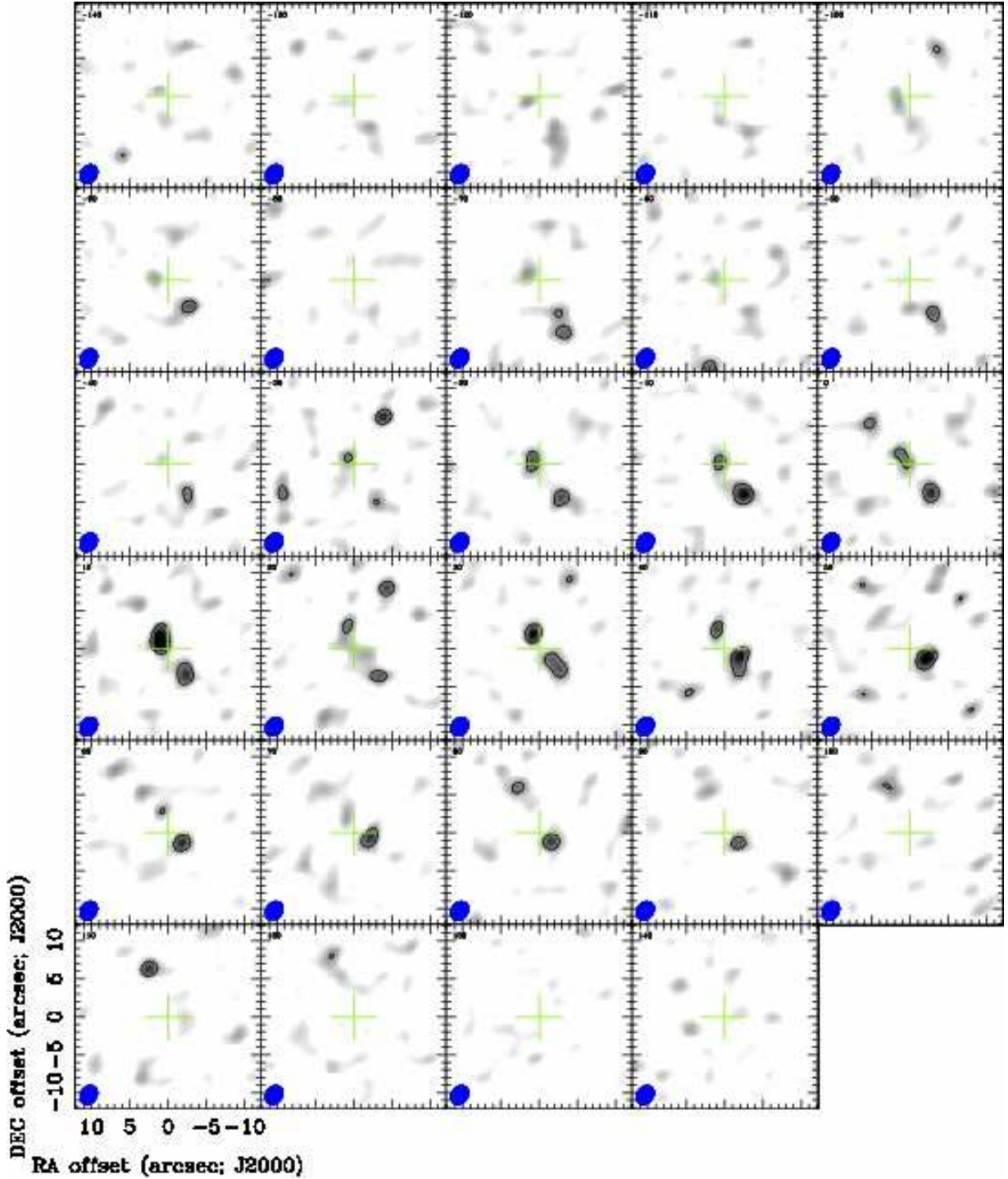}}
\caption{CO(3--2) channel maps of the galaxy NGC\,4569. The velocity relative to the systemic velocity of the galaxy ($v_{\rm hel}$$=$-235\,km\,s$^{-1}$) is given at the top left of each map. The phase center indicated by the large cross is at the AGN position $\alpha_{J2000}$$=$$12^h 36^m 49.8^s$, $\delta_{J2000}$$=$$13^{\circ}09'46.3''$. The beam (represented at the bottom left of each map) is 2.6$''$$\times$2.1$''$ and the rms is 115\,mJy\,beam$^{-1}$. The only contour line overlaid corresponds to the 4\,$\sigma$ level. The maps are not corrected for the primary beam attenuation.}\label{fig:n4569chanmaps}
\end{figure*}

\begin{figure*}
\centering
\resizebox{16cm}{!}{\includegraphics[width=16cm]{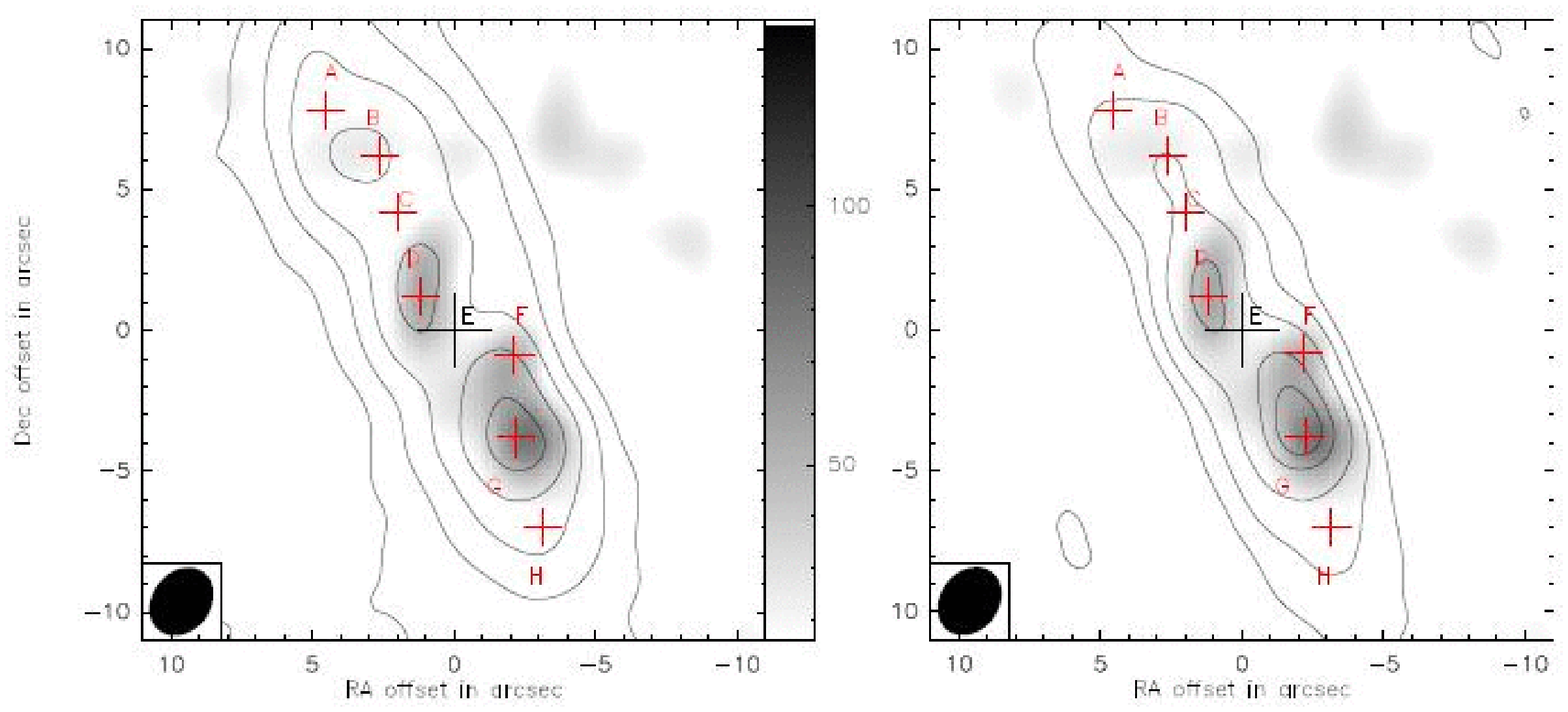}}

\caption{SMA CO(3--2) integrated map of NGC\,4569 (grey scale given in the color bar in Jy\,km\,s$^{-1}$\,beam$^{-1}$) overlaid with the PdBI  CO(1--0) integrated map (left) with contours at  2, 6, 12, 20, and 30 Jy\,km\,s$^{-1}$\,beam$^{-1}$ and the PdBI CO(2--1) integrated map (right) with contours at at  7, 21, 42, 70 and 87  Jy\,km\,s$^{-1}$\,beam$^{-1}$. No primary beam correction has been applied. All maps have the same resolution (PdBI data have been tapered) and the beam (2.3$''$$\times$2.1$''$) is shown at the bottom left. The black cross labelled E shows the position of the AGN ($\alpha_{\rm J2000}=12^{\rm h}36^{\rm m}49\ffs80$, $\delta_{\rm J2000}=13^{\circ}09'46\ffas3$). It is also the dynamical center and the phase center for the observations. The other crosses indicate the positions of the spectra plotted in Fig.\,\ref{fig:N4569spectra}}\label{fig:n4569mean}
\end{figure*}
\begin{figure}
\centering
\resizebox{8cm}{!}{
  \rotatebox{0}{
    \includegraphics{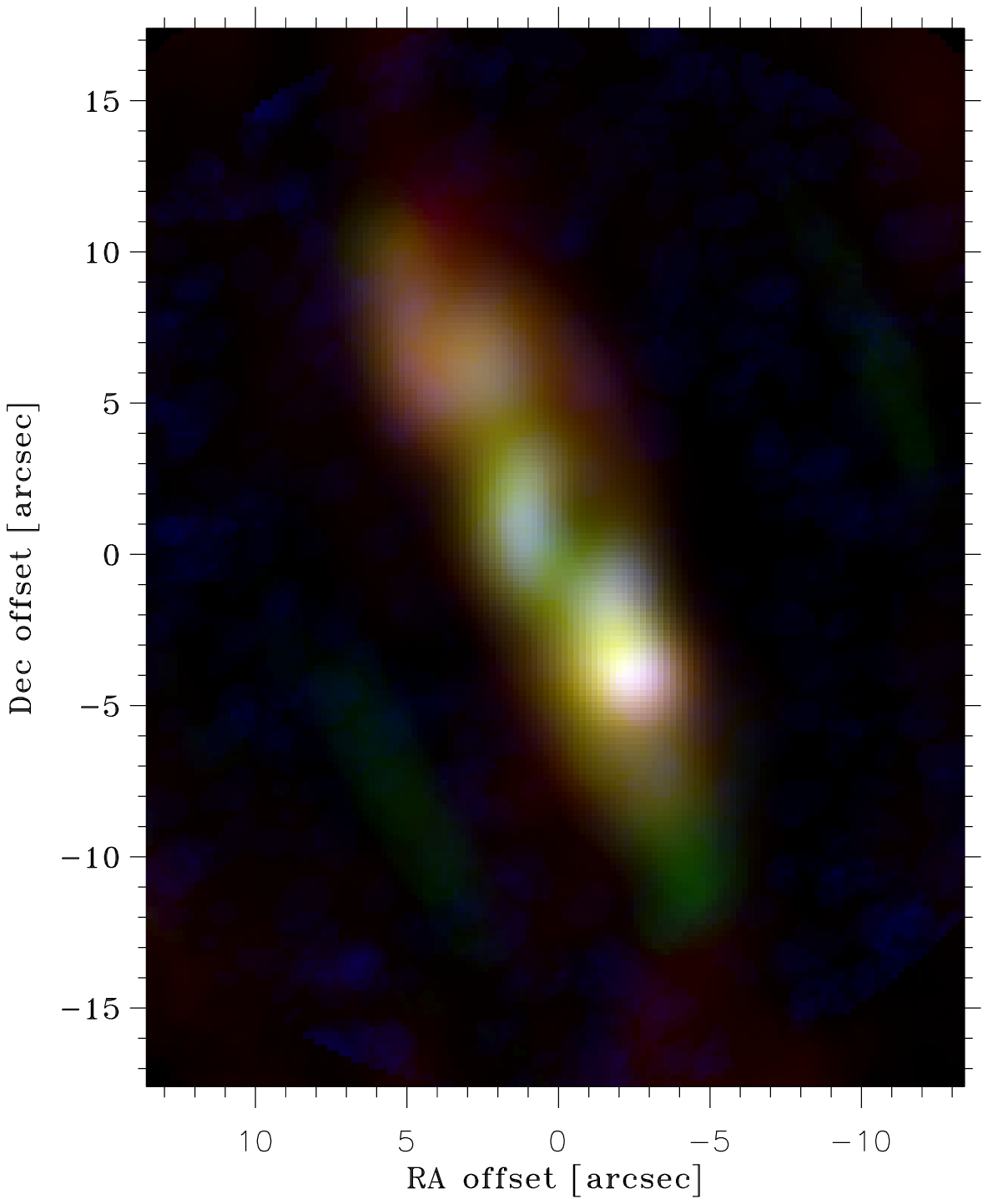}
  }
  
}

\caption{Composite color image of NGC\,4569 made from the intensity maps of the CO(1--0), CO(2--1) and CO(3--2) line emission. The line intensities are represented by the red, green and blue color intensities respectively. The intensity maps were obtained by summing all channels where the line has S/N$>$3, and are corrected for primary beam attenuation.}\label{fig:n4569rgb}
\end{figure}

\begin{figure*}
\centering
\resizebox{14cm}{!}{\includegraphics{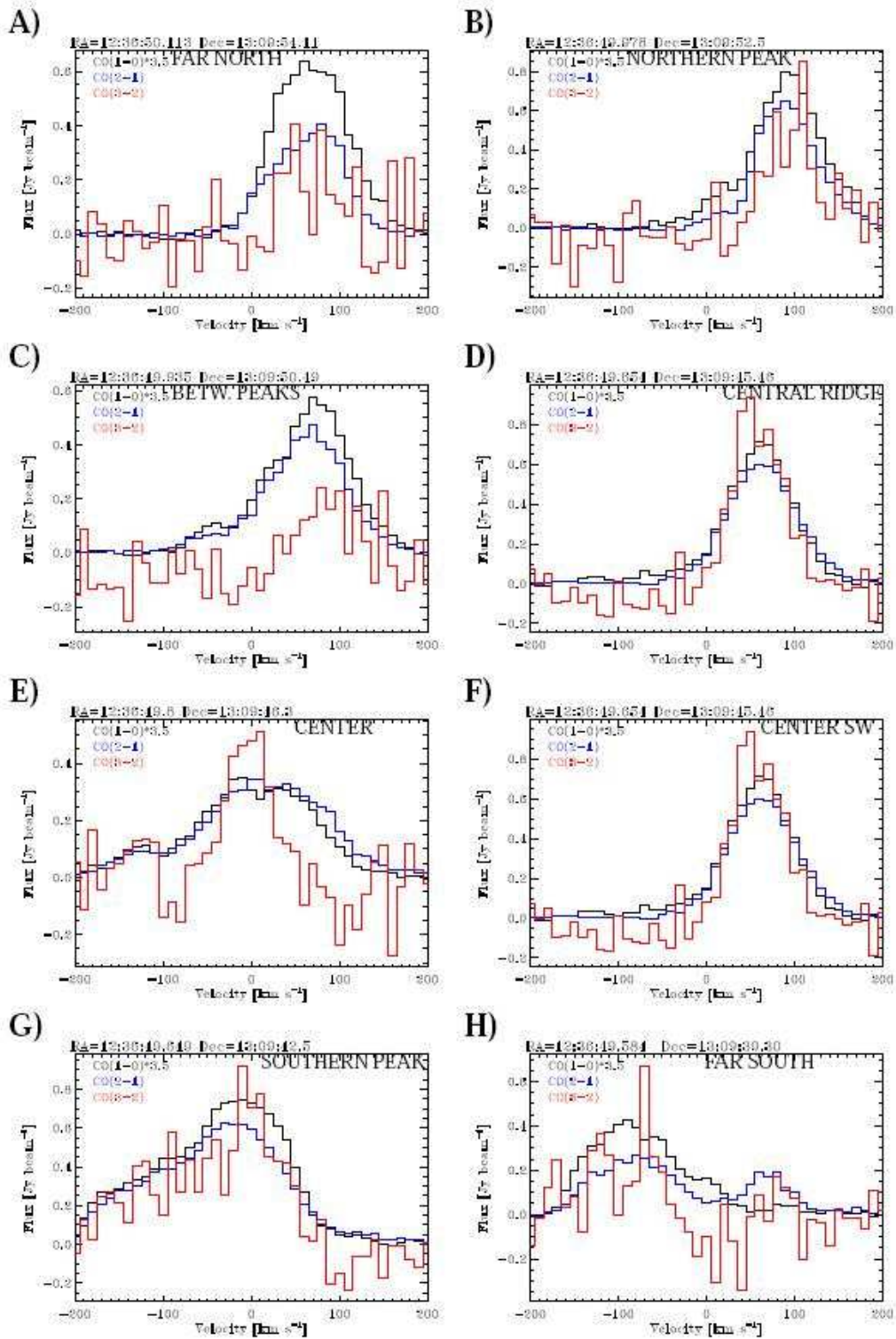}}

\caption{Individual spectra of CO(1--0) (black), (2--1) (blue) and (3--2) (red) lines at six different positions (see Fig.\,\ref{fig:n4569mean}) in the central region of NGC\,4569 extracted at the same spatial resolution of 2.6''$\times$2.1''. The fluxes have been corrected for primary beam attenuation and the CO(1--0) flux is scaled up by a factor 3.5 for better readability.}\label{fig:N4569spectra}
\end{figure*}

\begin{figure*}
\centering
\resizebox{16cm}{!}{\includegraphics[width=7cm]{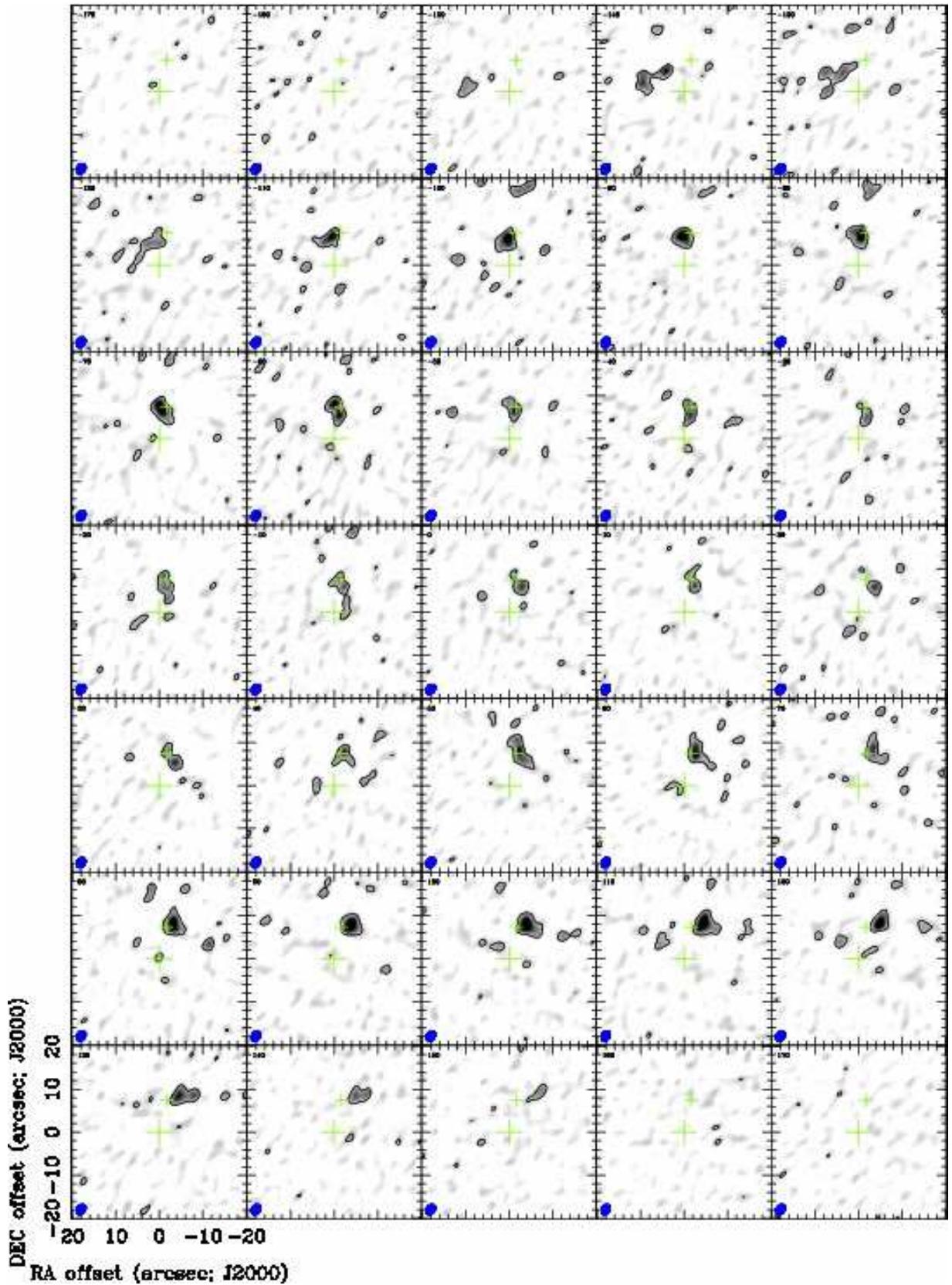}}
\caption{CO(3--2) channel maps of the galaxy NGC\,4826. The velocity relative to the systemic velocity of the galaxy ($v_{\rm hel}$$=$402\,km\,s$^{-1}$) is given at the top left of each map. The phase center indicated by the large cross is at $\alpha_{J2000}$$=$$12^h 56^m 43.76^s$, $\delta_{J2000}$$=$$21^{\circ}40'51.9''$. The galactic nucleus is marked by the small cross. The beam (represented at the bottom left of each map) is 2.6$''$$\times$1.9$''$ and the rms is 75\,mJy\,beam$^{-1}$. The only contour line overlaid corresponds to the 4\,$\sigma$ level.}\label{fig:n4826chanmaps}
\end{figure*}

\begin{figure*}
\centering
\centering
\resizebox{16cm}{!}{\includegraphics{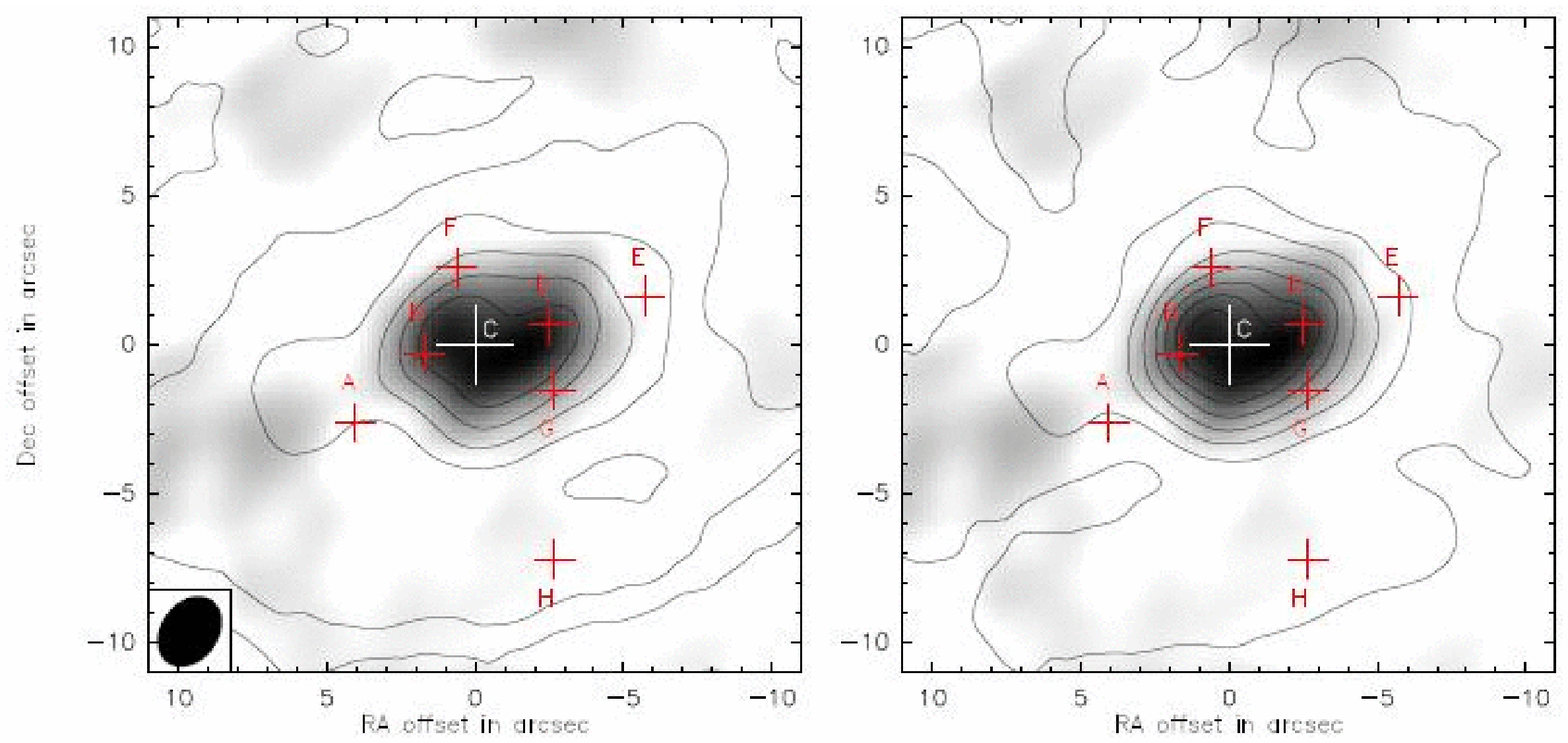}}

\caption{SMA CO(3--2) integrated map of NGC\,4826 (grey scale given in the color bar in Jy\,km\,s$^{-1}$\,beam$^{-1}$) overlaid with the PdBI  CO(1--0) integrated map (left) with contours at  2, 4, 8, 12, 16, 20, and 24 Jy\,km\,s$^{-1}$\,beam$^{-1}$ and the PdBI CO(2--1) integrated map (right) with contours at 4.1, 14.3, 24.5, 34.7, 44.9, 55.1, 65.3 and 75.5 Jy\,km\,s$^{-1}$\,beam$^{-1}$. No primary beam correction has been applied. All maps have the same resolution (PdBI data have been tapered) and the beam (2.6$''$$\times$1.9$''$) is shown at the bottom left. 
 The white cross labelled C shows the position of the dynamical center ($\alpha_{\rm J2000}=12^{\rm h}56^{\rm m}43\ffs64$, $\delta_{\rm J2000}=21^{\circ}40'59\ffas3$). The other crosses indicate the positions of the spectra plotted in Fig.\,\ref{fig:N4826spectra}}\label{fig:n4826mean}
\end{figure*}
\begin{figure*}
\centering
\resizebox{15cm}{!}{
  \rotatebox{0}{
    \includegraphics{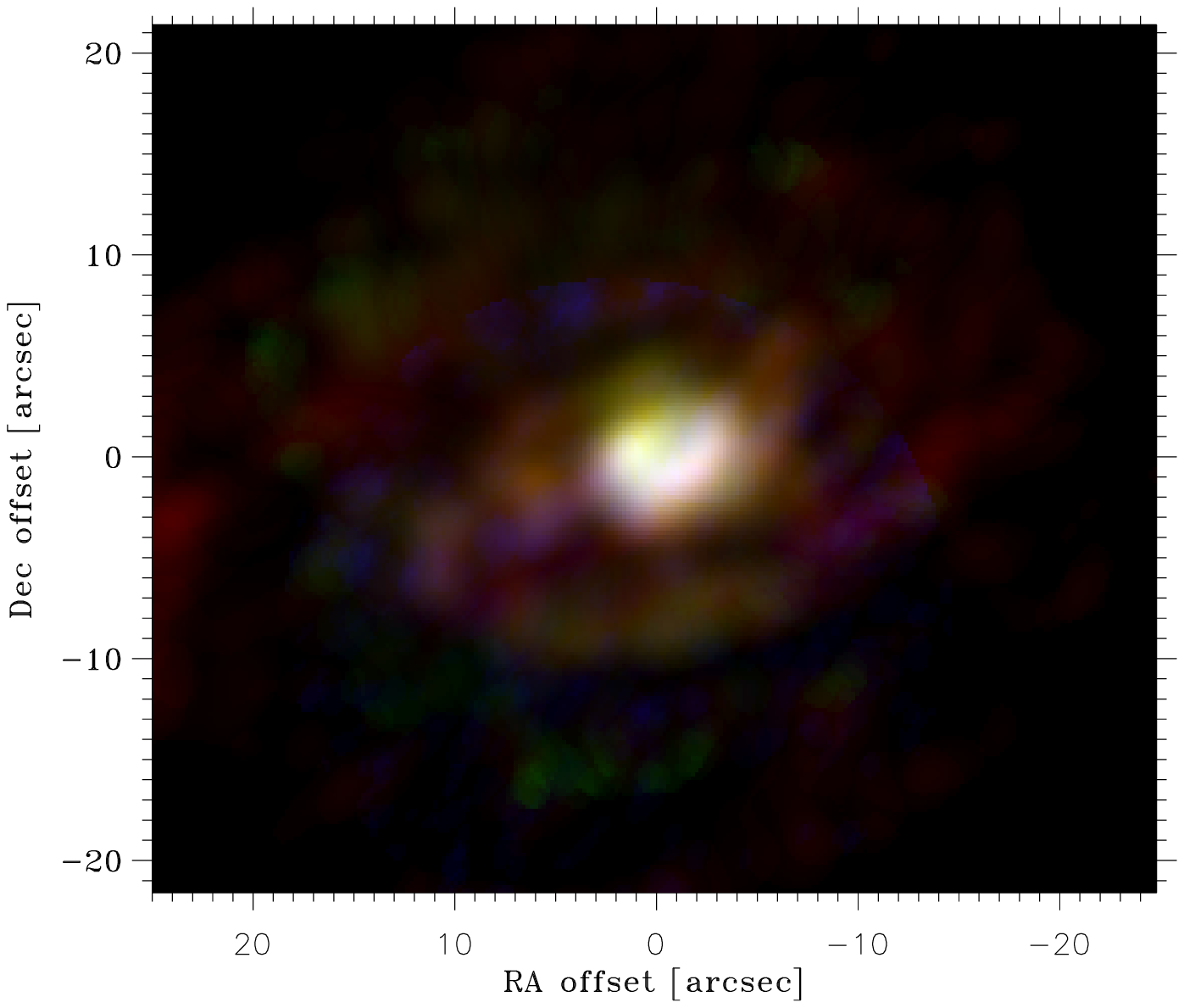}
  }
  
}

\caption{Composite color image of NGC\,4826 made from the intensity maps of the CO(1--0), CO(2--1) and CO(3--2) line emission. The line intensities are represented by the red, green and blue color intensities, respectively. The intensity maps were obtained by summing all channels where the line has S/N$>$3 and are corrected for primary beam attenuation.}\label{fig:n4826rgb}
\end{figure*}

\begin{figure*}
\centering

\resizebox{14cm}{!}{ \includegraphics{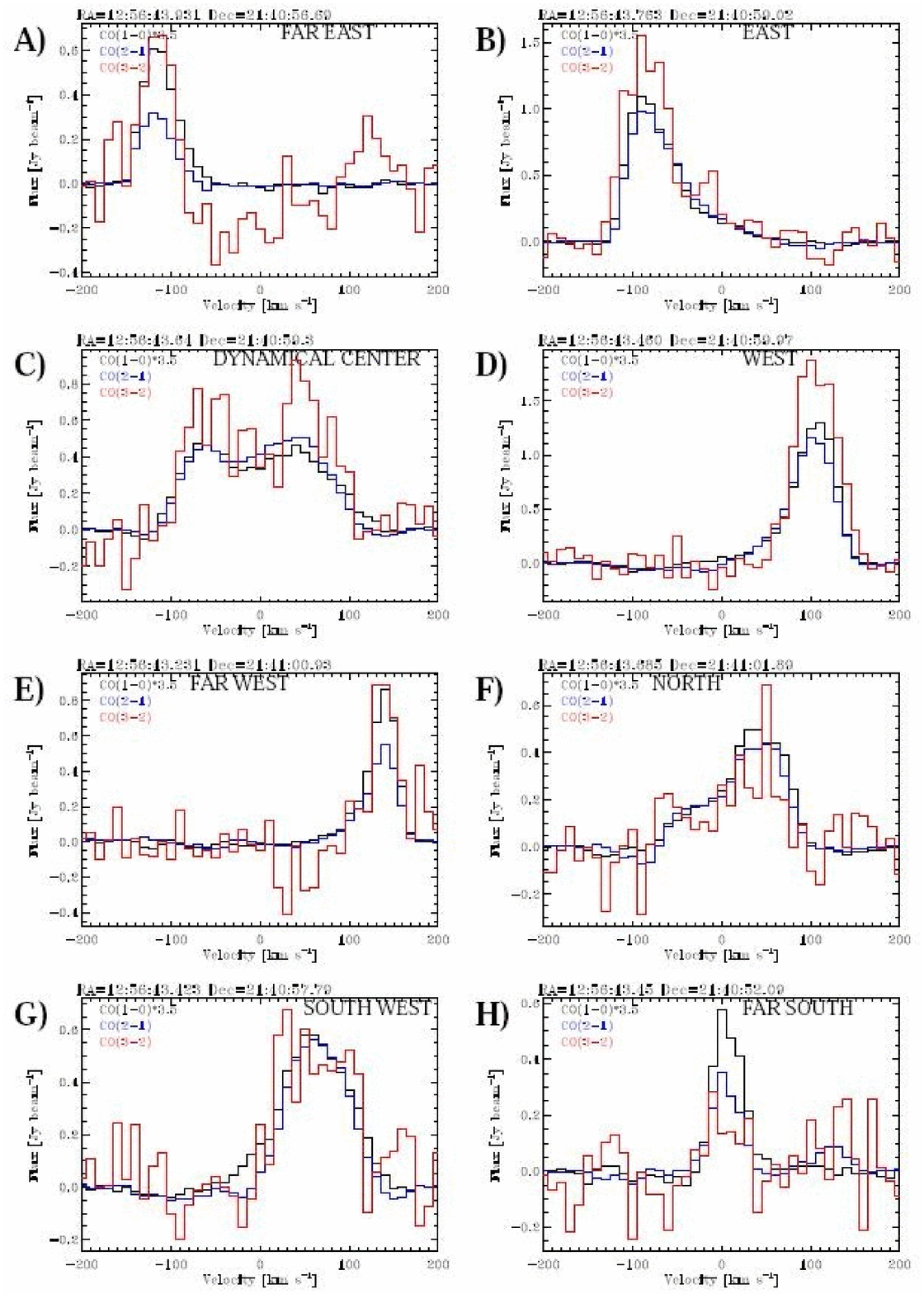}}

\caption{Individual spectra of the CO(1--0) line (black), the CO(2--1) line (blue) and the CO(3--2) line (red) at 8 different positions (see Fig.\,\ref{fig:n4826mean}) in the central region of NGC\,4826 extracted at the same spatial resolution of 2.6''$\times$1.9''. The fluxes have been corrected for primary beam attenuation and the CO(1--0) flux is scaled up by a factor 3.5 for better readability.}\label{fig:N4826spectra}
\end{figure*}


Each galaxy was observed with the SMA in an $\sim$8 hour track at night in March 2005. The observing frequency was set to the CO(3--2) line frequency  (345.796\,GHz) redshifted according to the systemic velocity of the galaxy. At these frequencies the half power beam width (HPBW) of the primary beam is $\sim$36$''$. The SMA was in its compact configuration with seven working antennas. Saturn was observed for bandpass calibration, bright nearby quasars (3C273 and 3C279 for NGC\,4569, 3C279 and  1159$+$292 for NGC\,4826) were observed  every $\sim$15\,min for gain calibration, and Titan and Callisto were used for flux calibration.
The data were reduced using the SMA MIR package \citep{mirref}. For image reconstruction the MIRIAD software\footnote{http://www.cfa.harvard.edu/sma/miriad/manuals/manuals.html} \citep{1995ASPC...77..433S} was used with natural weighting applied to the uv data.

We estimated the flux missed by the SMA due to the short spacings inherently missing in interferometric observations by comparing to the  single dish data published by \citet{2009ApJ...693.1736W} for NGC\,4569 and by \citet{2009A&A...493..525I} for NGC\,4826. We compared the total flux within an aperture of $\sim$36$''$ in diameter after correcting the SMA data for the primary beam attenuation and degrading to the resolution of the single dish data (14.5$''$).
 We found that 21\% and 38\% of the flux is missed by the SMA in NGC\,4569 and NGC\,4826, respectively.

We then used the single dish data to compute and add the missing short spacings to the SMA data. The corresponding large scale distribution was included in the image plane using the IMMERGE task of the MIRIAD software. One of the input parameters to this task is a factor to scale the flux of the single dish data. We controlled the value of this parameter by comparing the flux of the merged data with that of the single dish data following the same procedure as the one described above to estimate the missing flux. The factor was fixed such that the fluxes of the merged data and single dish data agreed to within 10\%. 

For NGC\,4569 the synthesized beam HPBW is  2.6$''$$\times$2.1$''$ with a position angle of $-$36\,deg and the RMS is 115\,mJy\,beam$^{-1}$ in 10\,km\,s$^{-1}$ channels. For NGC\,4826 the HPBW is 2.6$''$$\times$1.9$''$ with a position angle of $-$37\,deg and the RMS is 75\,mJy\,beam$^{-1}$ in 10\,km\,s$^{-1}$ channels.

To make the comparison between the three CO lines easier the PdBI observations (which also include the short spacings) have been degraded to the resolution of the SMA observations. This was done by apodizing the visibilities, cleaning the datacubes and restoring them with the same clean beam as the one used for the corresponding SMA map.



\section{Comparison of the three CO line intensity distributions}\label{sec:3lines}

\subsection{NGC\,4569}

The channel maps of the CO(3--2) line emitted by NGC\,4569 and observed with the SMA are presented in Fig.\,\ref{fig:n4569chanmaps}. The elongated distribution identified in most of the individual CO(1--0) and CO(2--1) channel maps  \citepalias[see][]{2007A&A...471..113B} can also be recognized in the CO(3--2) channel maps. The CO(3--2) emission appears more clumpy and irregular, however, and while emission was detected up to $\pm$200\,km\,s$^{-1}$ relative to systemic in  CO(1--0) and CO(2--1), there is no CO(3--2) emission at a level higher than 3\,$\sigma$ for velocities greater than 130\,km\,s$^{-1}$ in absolute value. These differences between the PdBI and SMA data may be due to the lower S/N ratio of the CO(3--2) SMA data. 
Indeed, the integrated map (see RGB image in Fig.\,\ref{fig:n4569rgb} and detailed grey scale maps in Fig.\,\ref{fig:n4569mean}) shows that  the  CO(3--2) is well detected at the peaks of CO(1--0) and CO(2--1) line emission, i.e. the northern (B) and the southern (G) peaks that were interpreted as emission from an elongated ring at the Inner Lindblad Resonance \citepalias[see][]{2007A&A...471..113B}, and at the central ridge (D). As in the two other line maps, there is a depression in the CO(3--2) map at the location of the nucleus (E) which is also the dynamical center, and the southern peak is more prominent than the northern one. 

There are, however, differences in the integrated maps as well. The depression at the center and the difference between the northern and southern peaks seem to be more pronounced in the CO(3--2) map. The CO(3--2) map also shows more emission west of the center (F) than the other line maps.

The spectra at the 8 positions marked by crosses in the integrated maps (Fig.\,\ref{fig:n4569mean}) and labelled from A to H are presented in Fig.\,\ref{fig:N4569spectra}. The SMA  spectra show negative values outside the CO(3--2) line. These are due to deconvolution errors (a single SMA track was obtained, whereas the PdBI observations include four tracks in four configurations, leading to better uv-coverage and therefore lower sidelobes). These effects are difficult to correct for without additional observations, but it can be seen that the negative 
offsets do not exceed 100\,mJy, i.e., the deconvolution errors do not exceed the RMS.

Although the CO(3--2) line has a lower S/N, the three CO lines look fairly similar in their profiles at a given position as well as in their spatial distribution, i.e., their intensity ratios are constant as a function of velocity to within $\sim$30\% over the profiles. This is particularly true at the three peaks (positions B, D and G). This implies that overall, the kinematics are the same for the three emission lines, i.e. even if different clumps or gas components are probed by the different lines they all seem to follow the same kinematics. The CO(1--0) emission seems to be relatively stronger away from the peaks (positions A, C and H); this is particularly striking  to the north (at position A), but a similar trend is seen toward the south in position H. 



\subsection{NGC\,4826}
The channel maps of the CO(3--2) emission from NGC\,4826 as observed with the SMA are presented in Fig.\,\ref{fig:n4826chanmaps}. The phase center is offset by $\sim$10$''$ to the south of the dynamical center, which is marked by a small cross on the channel maps. The typical disk rotation pattern  (a.k.a.\ spider diagram) clearly seen in  the individual CO(1--0) and CO(2--1) channel maps   \citepalias{2003A&A...407..485G} can just be recognized in the CO(3--2) channel maps. 
The integrated map (see RGB-image in Fig.\,\ref{fig:n4826rgb} and detailed greyscale maps in Fig.\,\ref{fig:n4826mean}), however, shows that  the  CO(3--2) distribution is similar to that of the CO(1--0) and CO(2--1) lines at the same resolution. It is dominated by a nearly circular region elongated along the galactic major axis with a radius of $\sim$2$''$ (40\,pc) and identified as a circumnuclear disk (CND) in  \citetalias{2003A&A...407..485G}. The emission detected outside the disk seems to coincide with the spiral arms described in  that same paper. This general agreement between the three CO lines suggests that the discrepancies in the channel maps are mainly due to the lower S/N of the SMA data in individual channels. This interpretation is further supported by the similarities between the individual spectra shown in Fig.\,\ref{fig:N4826spectra}. At the eight positions (marked on the integrated map in Fig.\,\ref{fig:n4826mean}) the spectra exhibit similar profiles to within 30\%.

The relative strengths of the lines appear to follow two trends. ($i$) The closer to the center, the stronger the CO(2--1) and CO(3--2) lines  relative to  CO(1--0). ($ii$) The CO(3--2) line is stronger relative to   CO(2--1) in the central disk (positions B, C and D) and along the major axis (compare positions A and E vs  F and G), where the spiral arms connect to the disk.

\section{Line ratios}\label{sec:lineratios}

\begin{figure*}
\centering
\resizebox{17cm}{!}{\includegraphics{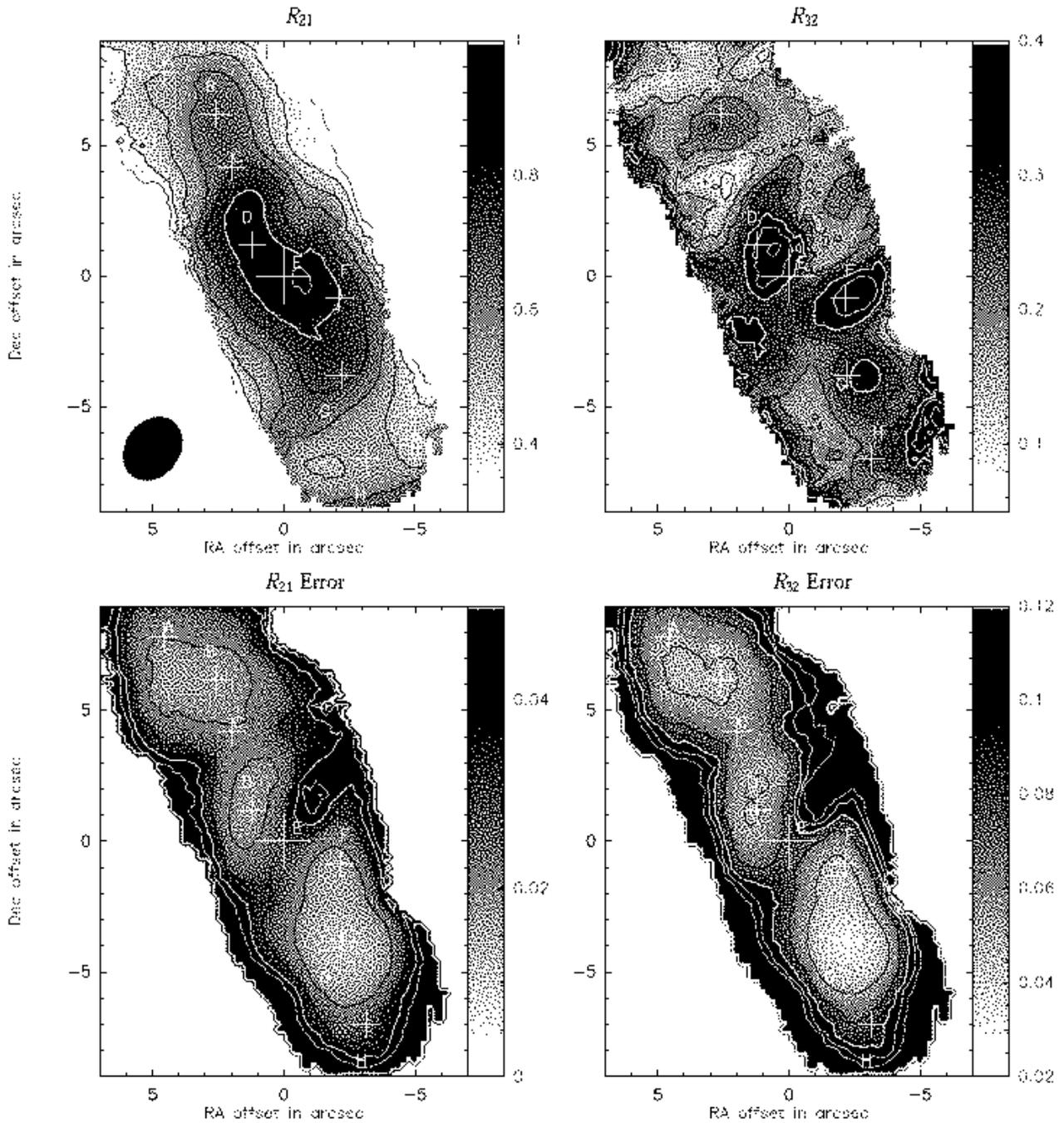}}

\caption{Line ratio maps of NGC\,4569 and associated uncertainties. 
{\bf Upper Left}: $R_{21}$ with black contours  ranging from 0.3 to 0.7 in steps of 0.1 and white contours at 0.8 and 0.9. The beam (2.3$''$$\times$2.1$''$) is represented as a filled ellipse at the bottom left corner.
{\bf Upper Right}: $R_{32}$ with black contours  ranging from  0.1 to 0.25 in steps of 0.05, and white contours at 0.3, 0.35 and 0.4.
{\bf Lower Left}: $R_{21}$ error map with black contours at 0.02, 0.03, and white contours at 0.04, 0.05. 
{\bf Lower Right}: $R_{32}$ error map with black contours 0.05, 0.07, and white contours at 0.09, 0.11, 0.13.
}\label{fig:n4569ratiomaps}
\end{figure*}

\begin{figure*}
\centering
\resizebox{17cm}{!}{\includegraphics{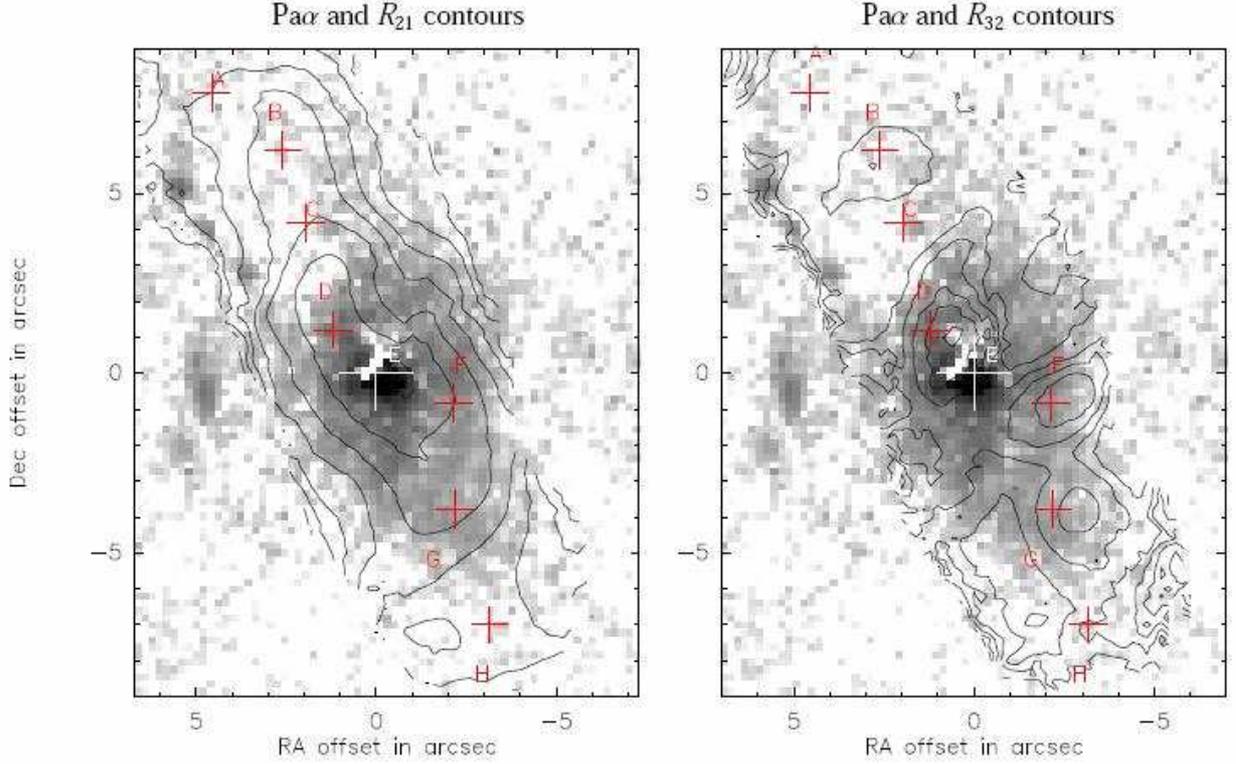}}

\caption{Line ratio maps with the same contours as in  Fig.\,\ref{fig:n4569ratiomaps} (only above 0.2 for $R_{32}$) overlaid on the  Pa$\alpha$ map obtained from NICMOS data (Programme ID: 9360, PI: Kennicutt). The grey scale is logarithmic; dark color corresponds to high flux.}\label{fig:n4569ratiomapsoverlays}
\end{figure*}

\begin{figure*}
\centering
\resizebox{19cm}{!}{\includegraphics{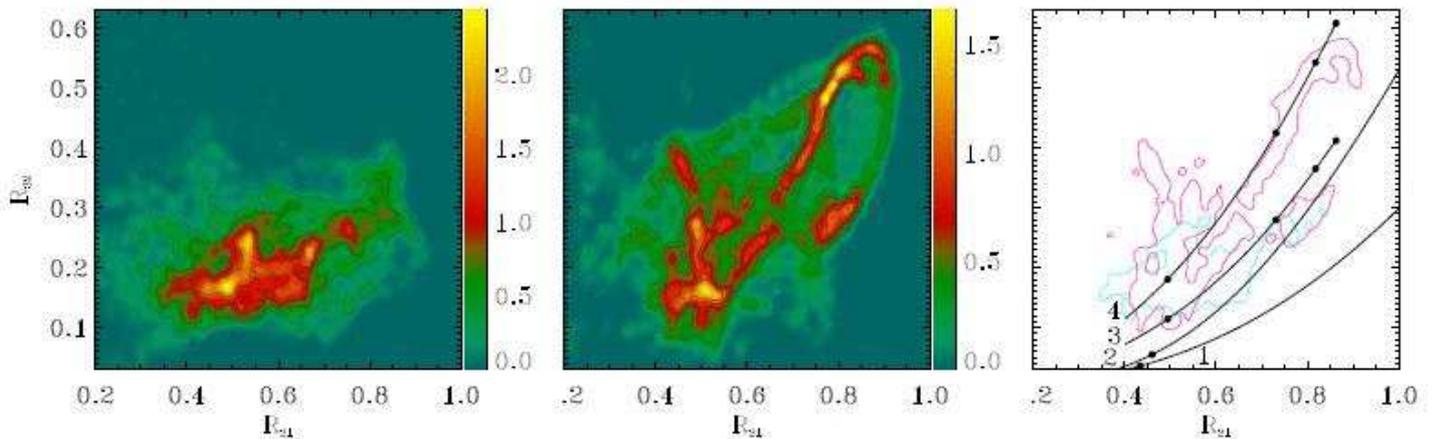}}
\caption{{\bf(Left)} $R_{32}$-vs-$R_{21}$ diagram for the galaxy NGC\,4569. The x-axis represents $R_{21}$ and the y-axis $R_{32}$.  The intensity scale gives the sky area in the map per unit area in the ratio space, it is expressed in beams per $0.01\times 0.01$ ratio area.  This diagram has been produced by convolving the actual distribution of pixels in the ratio-ratio plane by a gaussian kernel with a FWHM of 0.02. {Note that the pixels in the ratio maps used to compute the distribution are much smaller than the beam} (see Section \ref{sec:lineratios}). The contour levels are 0.3, 0.6, 0.9, 1.2 and 1.5 beams$/0.01^2$. {\bf (Middle)} $R_{32}$-vs-$R_{21}$ diagram for the galaxy NGC\,4826; the contour levels are 0.3, 0.5, 0.7, 0.9, 1.1 beams$/0.01^2$. 
{\bf (Right) } LTE and pseudo-LTE models represented as black curves in the $R_{32}$-vs-$R_{21}$ diagram.
Curves 1 and 2 correspond to pure LTE models with $N_{\rm CO}/dV=10^{15}$\,cm$^{-2}$\,pc\,(km\,s$^{-1}$)$^{-1}$ and $N_{\rm CO}/dV=12\times 10^{15}$\,cm$^{-2}$\,pc\,(km\,s$^{-1}$)$^{-1}$, respectively. The temperature increases from 4\,K at the bottom left to 8\,K at the top right.
Curves 3 and 4 correspond to pseudo-LTE models with $f_{\rm 32}=0.5$ and $f_{\rm 32}=0.2$ respectively (see Section\,\ref{sec:lte} for the definition of $f_{\rm 32}$). The column density to line width ratio is $N_{\rm CO}/dV=10^{15}$\,cm$^{-2}$\,pc\,(km\,s$^{-1}$)$^{-1}$ and the temperature increases from 4 to 20\,K along the curves from bottom left to top right. The cyan and magenta curves show the lowest contours for NGC\,4569 and NGC\,4826, respectively.
}\label{fig:ratioratiolte}
\end{figure*}

\begin{figure*}
\centering
\resizebox{17cm}{!}{\includegraphics{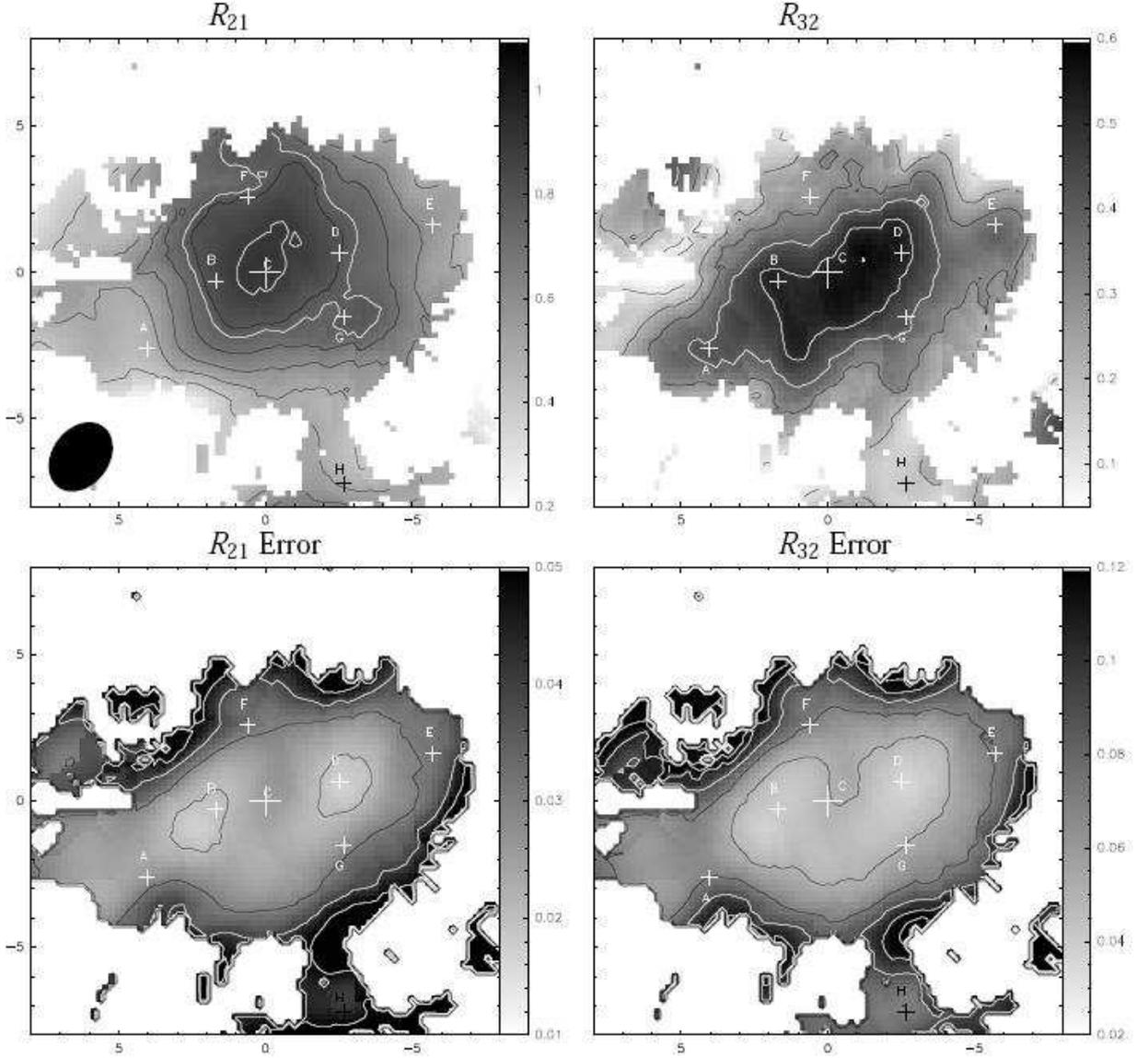}}

\caption{Line ratio maps of NGC\,4826 and associated uncertainties. The crosses are the same as in Fig.\,\ref{fig:n4826mean}.
{\bf Upper left}: $R_{21}$ with black contours at 0.4, 0.5, 0.6, 0.7, 0.8 and white contours at 0.77 and 0.9.  The beam (2.6$''$$\times$1.9$''$) is represented as a filled ellipse at the bottom left corner.
{\bf Upper right}: $R_{32}$ with black contours at 0.1, 0.2, 0.3 and white contours at 0.4 and 0.5. 
{\bf Lower left}: $R_{21}$ error map with black contours at 0.02, 0.03 and white contours at 0.04 and 0.05. 
{\bf Lower right}: $R_{32}$ error map with black contours at 0.05, 0.07 and white contours at 0.09 and 0.11 
}\label{fig:n4826ratiomaps}
\end{figure*}

\begin{figure*}
\centering
\resizebox{17cm}{!}{\includegraphics{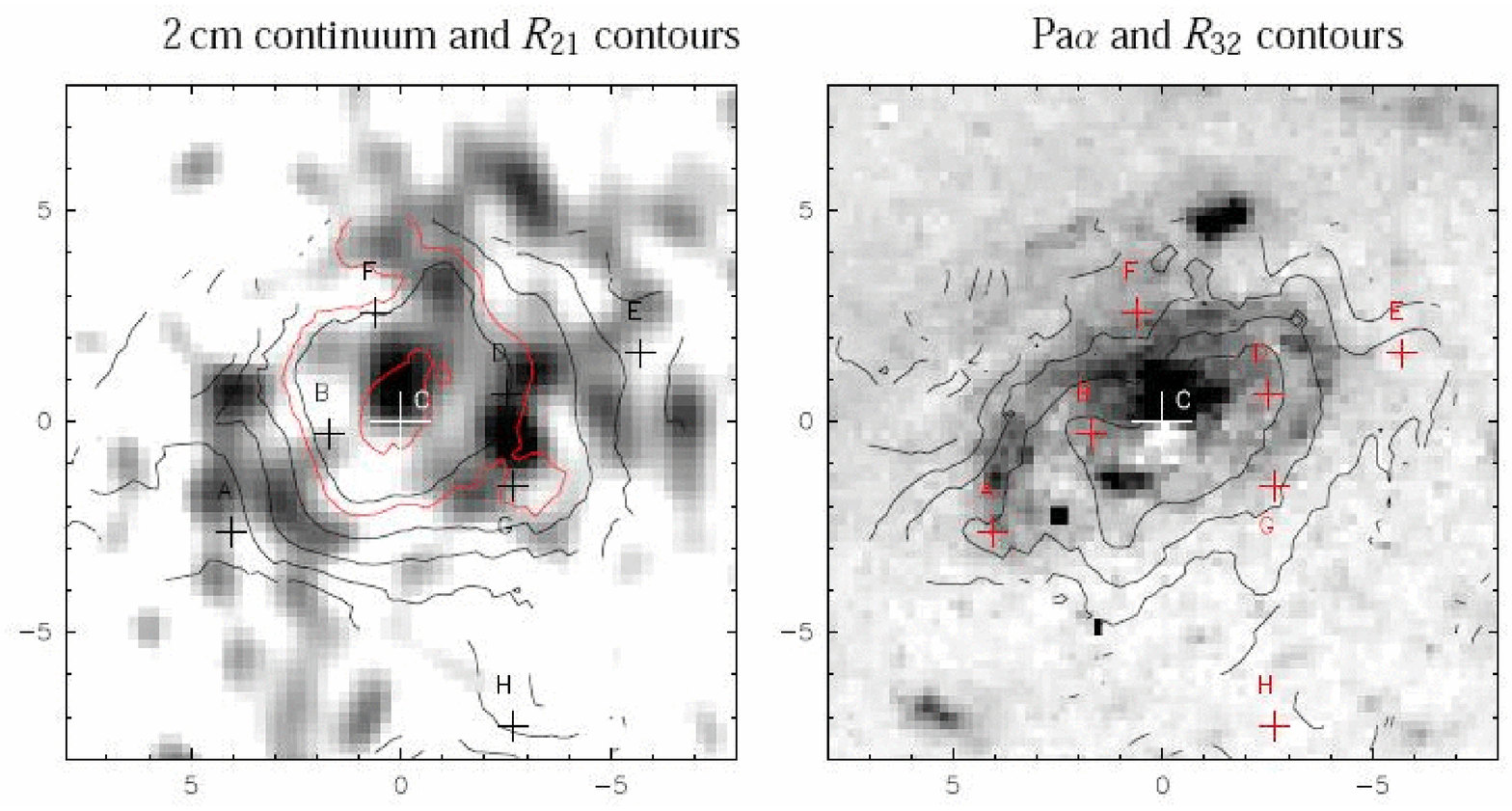}}

\caption{Maps of NGC\,4826. {\bf Left}: $R_{21}$ with the same contours as in Fig.\,\ref{fig:n4826ratiomaps} overlaid on the 2cm radio continuum map from \citet{1994ApJ...421..122T}. 
{\bf Right}: $R_{32}$ with the same contours as in Fig.\,\ref{fig:n4826ratiomaps} overlaid on the Pa$\alpha$ map from \citetalias{2003A&A...407..485G}. Dark color corresponds to high flux.  
}\label{fig:n4826ratiomapsoverlays}
\end{figure*}

 With the definition adopted in the introduction any line ratio would tend to 1 with increasing excitation temperature for thermalized gas in local thermal equilibrium.
Line ratio maps were computed restricting the spectral window of each pixel to the channels where the CO(1--0) flux is $>$4-$\sigma$ and the CO(3--2) flux is positive. When the spectral window was narrower than 50\,km\,s$^{-1}$, i.e. less than 5 channels, the pixel was discarded. 

\subsection{NGC\,4569}
The ratio maps with the corresponding error (standard deviation) maps are shown in Fig.\,\ref{fig:n4569ratiomaps}. { $R_{21}$ is distributed in the range 0.2 to 0.9 with a weighted mean of 0.63. The uncertainties on  $R_{21}$ are in the range 0.01 to 0.05. $R_{32}$ is distributed in the range 0.05 to 0.45 with a weighted mean of 0.23. The uncertainties on  $R_{32}$ are in the range 0.03 to 0.11. The errors on the weighted means based on the data noise only are less than 0.02. While calibration uncertainties introduce an overall error of the order of 0.05, they should have the same effect on all locations and thereby spatial trends should be unaffected.
The mean values are in good agreement with the values found by \citet{2003A&A...398..959H} based on single dish observations with a beam of 80$''$ HPBW, i.e. $R_{21}$=0.64$\pm$0.08 and $R_{32}$$=$0.23$\pm$0.04; and with the $R_{32}$ value found by \citet{2009ApJ...693.1736W} within the inner 22$''$, i.e., 0.34$\pm$0.11 (note however that the latter is not a fully independent check as we used the same CO(3--2) single dish data to complement our SMA data).  }
 
 The $R_{21}$ and  $R_{32}$ maps are different. While $R_{21}$ is smooth, more or less symmetric with respect to the center, and continuously increasing toward the center with a maximum at $\sim$1$''$ to the west of the galactic center,  $R_{32}$ is more clumpy, exhibiting several peaks, and is more asymmetric. { The large-scale gradient in  $R_{32}$ reported by \citet{2009ApJ...693.1736W}  does not appear in the inner 10$''$}.\\

 $R_{32}$ is expected to probe more extreme conditions such as those occuring in star forming or shock regions. This is confirmed by the comparison to the Pa$\alpha$ emission (Figure\,\ref{fig:n4569ratiomapsoverlays}). Although the detailed structure of $R_{32}$ is not reflected in  Pa$\alpha$, most of the pixels where $R_{32}$$>$0.2 coincide with Pa$\alpha$ emission detected in the NICMOS image. $R_{21}$, instead, reaches high values outside (to the north) as well as inside the Pa$\alpha$ emission region, although it peaks at the galactic center as the Pa$\alpha$. The fact that $R_{32}$ tracks massive star formation in NGC\,4569 better than $R_{21}$ is consistent with the conclusion of \citet{2009ApJ...693.1736W}, based on maps over larger areas, that CO(3--2) is a robust tracer of star formation.

 The fact that $R_{32}$ correlates with the individual CO line intensities (the three peaks in CO are also peaks in  $R_{32}$) also suggests that it is more sensitive to density. In contrast, the fact that $R_{21}$ does not show the same structures as the individual line maps and that it increases steadily towards the center suggests that it is more sensitive to temperature. This is expected for thermalized gas with moderate opacity in the CO(3--2) line. 
  
 The left panel of Fig.\,\ref{fig:ratioratiolte} shows the distribution of the ratios in the $R_{32}$  vs $R_{21}$  space; in the following this will be referred to as the $R_{32}$-vs-$R_{21}$ diagram. { This diagram was obtained by mapping each pixel  (0.2$''$$\times$0.2$''$ in size) of the ratio maps  to the grid and convolving with a gaussian kernel of 0.02 FWHM,
corresponding to a typical uncertainty in the line ratio. Based on the line ratio error maps shown in Fig.\,\ref{fig:n4569ratiomaps} it should be kept in mind that any feature in this diagram with a size lower than $\sim$0.05 cannot be considered as real}.   { In addition, although the ratio maps are considered as continuous distributions (we used small pixels with respect to the beam), it should  be kept in mind that the resolution is finite in the ratio maps.}   {This diagram is used to show in a synoptic way which regions of the ratio-ratio space are populated. }
As expected from the differences between the $R_{32}$  and $R_{21}$ maps noted above, the distribution is dispersed in the diagram. The ratios populate a region going from (0.4, 0.15) to (0.8, 0.25) with an average thickness of $\sim$0.15 in an orthogonal direction.

\subsection{NGC\,4826}\label{sec:lineratios_n4826}

The ratio maps with the corresponding error (standard deviation) maps are shown in Fig.\,\ref{fig:n4826ratiomaps}. { $R_{21}$ is distributed in the range 0.2 to 0.95 with a weighted mean of 0.67.  $R_{32}$ is distributed in the range 0.05 to 0.60 with a weighted mean of 0.38. The errors on the ratios and the weighted means are similar to those in the case of NGC\,4569. 
The mean values are lower than the ratios found by \citet{2009A&A...493..525I} based on single dish observations in the inner 21$''$, i.e. $R_{21}$$=$0.98$\pm$0.17 and $R_{32}$$=$0.59$\pm$0.09 and the inner 12$''$, i.e. $R_{21}$$=$1.06$\pm$0.22 and $R_{32}$$=$0.79$\pm$0.18. They are, however, within (or close to) the 2$\sigma$ interval of the single dish measurements. }
 $R_{32}$ does not appear to be more clumpy than  $R_{21}$, in contrast to NGC\,4569, and both maps look relatively symmetric with respect to the dynamical center, with both ratios increasing steadily toward the center. There are however differences. While the $R_{21}$ spatial distribution appears to be close to circularly symmetric, the $R_{32}$ distribution is more elongated along the galactic major axis. Furthermore, $R_{21}$ seems to be more prominent to the north of the dynamical center (although its peak coincides with the dynamical center), whereas $R_{32}$ forms a ridge along the major axis south of the dynamical center with its peak $\sim$1$''$ to the southwest. 

As in  NGC\,4569, high $R_{32}$ values ($>0.3$) occur within the Pa$\alpha$ emission region as detected by NICMOS, while this is not necessarily true for $R_{21}$ (see Fig.\,\ref{fig:n4826ratiomapsoverlays} right panel). Indeed, although $R_{21}$ increases toward the center as $R_{32}$, it reaches high values ($>$0.7) north and south of the Pa$\alpha$ emission region, which is elongated along the galactic major axis.   
$R_{21}$ shows some correlations with the 2\,cm radio continuum (see Fig.\,\ref{fig:n4826ratiomapsoverlays} left panel). 
In particular, the 2\,cm peaks 1$''$ and 4$''$ north of the dynamical center coincide with high $R_{21}$ values. 
At the VLA's resolution, the 2\,cm continuum is dominated by 
emission from HII regions and SNRs around which the molecular phase is expected to be strongly heated. If $R_{21}$ is more sensitive to temperature relative to $R_{32}$, this may explain the better correlation of  $R_{21}$ with the 2\,cm continuum than $R_{32}$. 
We emphasize, however, that this trend is weak relative to the radial increase toward the center of $R_{21}$, $R_{32}$  and Pa$\alpha$ emission { \citep[as for $R_{21}/R_{32}$ , previously noted by ][]{2009A&A...493..525I}}. In fact both line ratios are sensitive to temperature and density and their interpretation in terms of physical conditions requires radiative transfer modeling;  this is the topic of the next section.

The middle panel of Fig.\,\ref{fig:ratioratiolte} shows the $R_{32}$-vs-$R_{21}$ diagram for NGC\,4826.  
The emission is much more widely spread in the $R_{32}$-vs-$R_{21}$ diagram than for NGC\,4569. The top part of the diagram with $R_{32}>0.3$ being more populated, in particular the top right part with $R_{21}>0.6$ and $R_{32}>0.3$. 
 
\section{Modeling the physical conditions of the molecular gas}\label{sec:physcond}

\begin{figure*}
\centering
\resizebox{18cm}{!}{\includegraphics{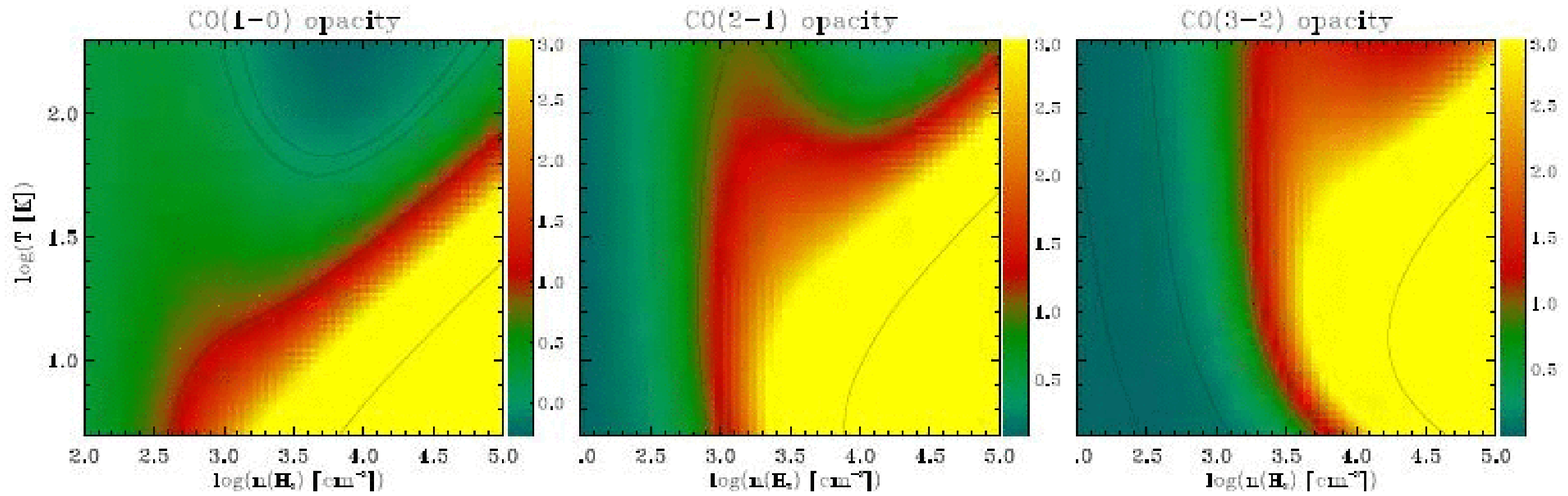}}
  
\caption{Variations of the CO line opacities  with respect to temperature and density based on LVG modeling.
From left to right panel: CO(1--0), (2--1) and (3--2) opacities. Used for the qualitative interpretation of the pseudo-LTE model.}\label{fig:tau}
\end{figure*}

\subsection{Pseudo-LTE model}\label{sec:lte}

The line intensity of a single component medium in local thermal equilibrium (LTE) is entirely determined by the temperature of the medium, $T$, and by the ratio of the column density to the local line width, $N_{\rm CO}/dV$ when the line is not optically thick. In reality, the ISM is multi-component and the intensity of each line  results from the sum over the beam of all the individual components' line intensities. LTE may still be a reasonable approximation for the local emission of individual components or clumps, but the average opacity in a line toward the observer also depends on more global properties of the ISM (e.g. turbulence and complex structures). A way to improve the model is to decouple the average opacity from the average local physical conditions and consider a mixture of components.   For the sake of simplicity we assume that CO(1--0) and CO(2--1) emissions are optically thick, and we consider optically thick ($\tau$$>>$1) and optically thin ($\tau$$<<$1) components for the CO(3--2) emission. This choice is justified by the variations of the line opacities with the physical conditions as shown in Fig.\,\ref{fig:tau} and based on large velocity gradient model (LVG model described in Section\,\ref{sec:lvg}). For gas with a temperature close to or below $\sim 20$\,K and a density above $\sim 10^3$\,cm$^{-3}$, the CO(1--0) and (2--1) emission is mainly optically thick whereas the CO(3--2) emission optical depth can vary from thick to thin. 
We denote $f_{32}$ the fraction of optically thin emission for CO(3--2). 
Treating $f_{32}$ as an independent parameter allows us to avoid complex modeling of the ISM, but it should be kept in mind that this parameter is also related to the physics of the ISM. In other words, the interpretation of such a model requires further physical modeling (at least qualitative) of the opacities. We assume the line intensities are given by the LTE equations { (the optically thick component is independent of $N_{\rm CO}/dV$)}. 
We refer to this modified LTE model as a pseudo-LTE model. { Although this model relies on simple assumptions, we think it can be used to interpret the overall distribution of the gas in the $R_{32}$-vs-$R_{21}$ diagram in terms of variations of its physical properties. 
In practice, we reproduce the overall distribution in the $R_{32}$-vs-$R_{21}$ plot with
families of pseudo-LTE models corresponding to different combinations of
physical parameters. In other words, this model is not used to fit each  ($R_{21}$, $R_{32}$) pair independently (this would not be possible given the number of unknowns); it is used to relate the different parts of the $R_{32}$-vs-$R_{21}$ diagram to the physical conditions and reproduce the overall ratio distribution.}

\subsubsection{NGC\,4569}

As shown in the right panel of Fig.\,\ref{fig:ratioratiolte}, a single-component LTE model (curves 1 and 2) cannot reproduce the line ratios observed in NGC\,4569 (cyan contour). Even with high column densities per unit velocity (curve 1 corresponds to $N_{\rm CO}/dV=10^{15}$\,cm$^{-2}$\,(km\,s$^{-1}$)$^{-1}$ and curve 2 to $N_{\rm CO}/dV=12\times 10^{15}$\,cm$^{-2}$\,(km\,s$^{-1}$)$^{-1}$) at the low-temperature (lower left) end  $R_{32}$ is always lower than observed.

 Curves 3 and 4 correspond to pseudo-LTE models with $f_{32}$$=$$0.5$ and  $f_{32}$$=$$0.2$ respectively. The column density to line width ratio is $N_{\rm CO}/dV=10^{15}$\,cm$^{-2}$\,(km\,s$^{-1}$)$^{-1}$ and the temperature goes from 4 to 20\,K along the curves from bottom left to top right.   Neither curve can fit the overall ratio distribution, although each one intersects with the distribution. The ratios corresponding to the bottom left of the diagram correspond to gas at $\sim$5\,K with the three lines being  mainly optically thick (curve 4) and the ridge at the top right of the distribution to gas at $\sim$15\,K with 50\% of the CO(3--2) line emitted by optically thin gas (curve 3). Thus, the overall distribution of the ratios in the $R_{32}$-vs-$R_{21}$ diagram  noted in the previous section could correspond to gas spanning the whole range of physical properties between these two limits, i.e. ($T$$=$5, $f_{32}$$=$0.2) and ($T$$=$15\,K, $f_{32}$$=$0.5). For the gas to remain within the observed ratio distribution the CO(3--2) thin fraction, $f_{32}$,  must increase with the temperature. To qualitatively interpret the variations of the optically thin fractions we consider the opacities given by a large velocity gradient  model  for a range of temperatures and densities  and displayed in Fig.\,\ref{fig:tau}. Qualitatively, gas that is optically thick in the three lines is denser ($n>10^4$\,cm$^{-3}$) than gas that is optically thin in the CO(3--2) line but still thick in the other two lines  ($10^3<n<10^4$\,cm$^{-3}$) for $T$$\le$20\,K. Hence, increasing  $f_{32}$ means increasing the fraction of more diffuse gas, and from left to right the ratio distribution (i.e. the distribution in the $R_{32}$-vs-$R_{21}$ diagram) corresponds to increasing temperature and decreasing average density.

\subsubsection{NGC\,4826}\label{sec:lte4826}
%
%

As shown in the right panel of Fig.\,\ref{fig:ratioratiolte}, The highest ratios observed in NGC\,4826 (magenta contour) follow more or less the pseudo-LTE model represented by the curve 4 in the $R_{32}$-vs-$R_{21}$ diagram. 
As noted in Section\,\ref{sec:lineratios_n4826} the line ratios increase toward the galactic center (Fig\,\ref{fig:n4826ratiomaps}), although their maxima do not exactly coincide with it. Hence according to this simple model the temperature should increase toward the center and  a significant fraction ($\sim$20$\%$) of the CO(3--2) emission should be optically thin.

\begin{figure*}
\centering
\resizebox{17cm}{!}{ \includegraphics{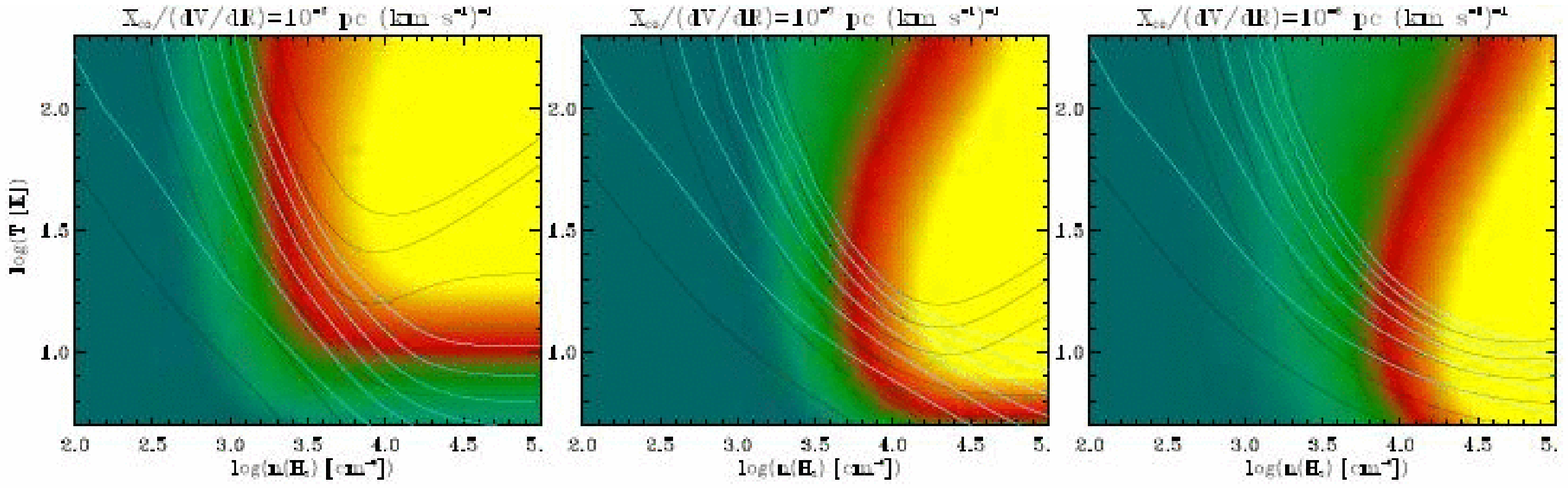}}

\caption{Grids of LVG models for a range of molecular hydrogen densities and kinetic temperatures. Each grid corresponds to a different value of  $X_{\rm CO}/(dV/dR)$, from left to right: $10^{-6}$, $10^{-7}$ and $10^{-8}$\,pc\,(km\,s$^{-1}$)$^{-1}$. The CO(1--0) radiative temperature is represented in color scale and is overlaid with $R_{21}$ black contours (0.3 to 1.3 in steps of 0.2) and  $R_{32}$ blue contours (0.1 to 0.6 in steps of 0.1). The line ratios increase from bottom left to top right.}\label{fig:lvggrid}
\end{figure*}

\begin{figure*}
\centering
\resizebox{17cm}{!}{\includegraphics{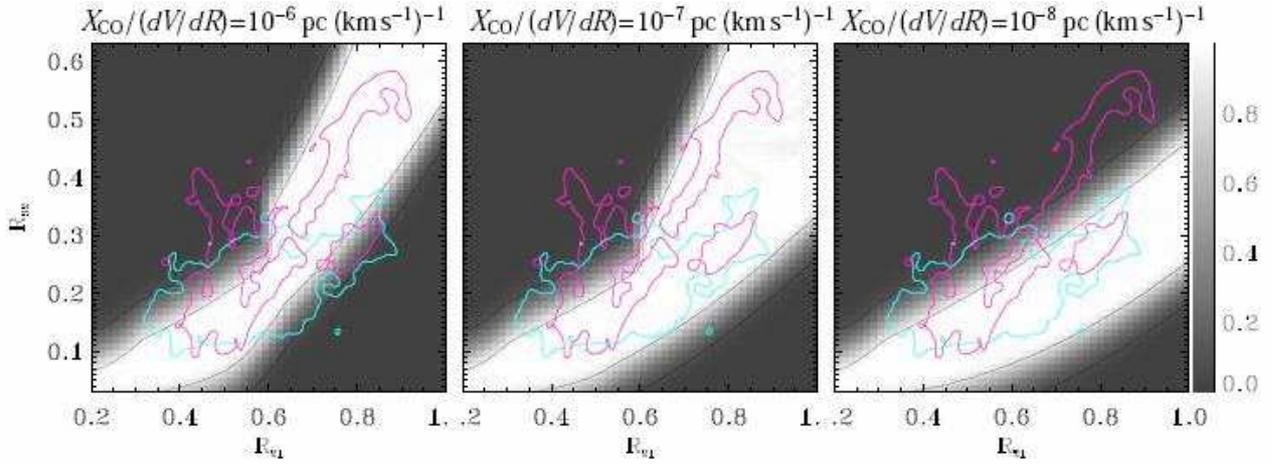}}

\caption{Likelihood of all possible ($R_{21}$,$R_{32}$) pairs on a $R_{32}$-vs-$R_{21}$ grid for different values of $X_{\rm CO}/(dV/dR)$. The contours of the $R_{32}$-vs-$R_{21}$ diagrams of NGC\,4569 (blue) and NGC\,4826 (magenta) are overlaid. The contours are smoothed and their levels correspond to 3 and 9 pixels per grid bin. (see Fig.\,\ref{fig:ratioratiolte}). }
\label{fig:gridgof}
\end{figure*}
\begin{figure*}
\centering
\resizebox{15cm}{!}{ \includegraphics{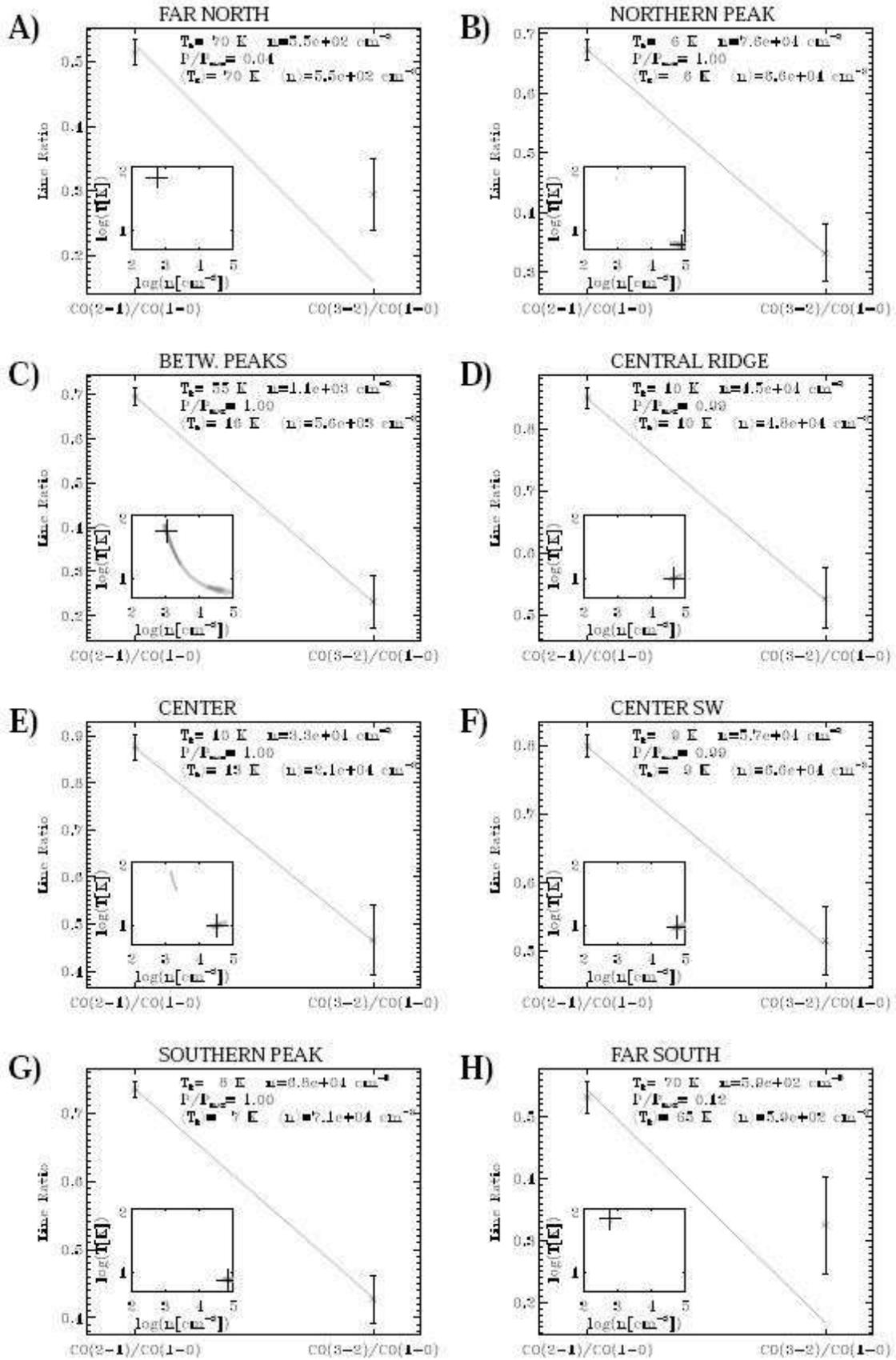}}

\caption{LVG best fits to the line ratios measured in the central region of NGC\,4569 at the eight positions marked in  Fig.\,\ref{fig:n4569mean}. The measured line ratios are represented with the 1-$\sigma$ error bars and the best fit model ratios are linked with a straight line. The inset in each plot shows the 90\% confidence interval (light grey shaded area) and the 50\% confidence interval (dark grey) in the density-temperature plane assuming a normal probability distribution for each ratio; the best fit solution is marked with a cross. The best fit temperature and density as well as the averages over the 90\% confidence interval are given at the top right of each plot. The normalized likelihood of the best fit, $P/P_{\rm max}$  is also given (1 means the measured values are exactly reproduced by the model).}\label{fig:N4569lvgfits}
\end{figure*}




\begin{figure*}
\centering
\resizebox{17cm}{!}{\includegraphics{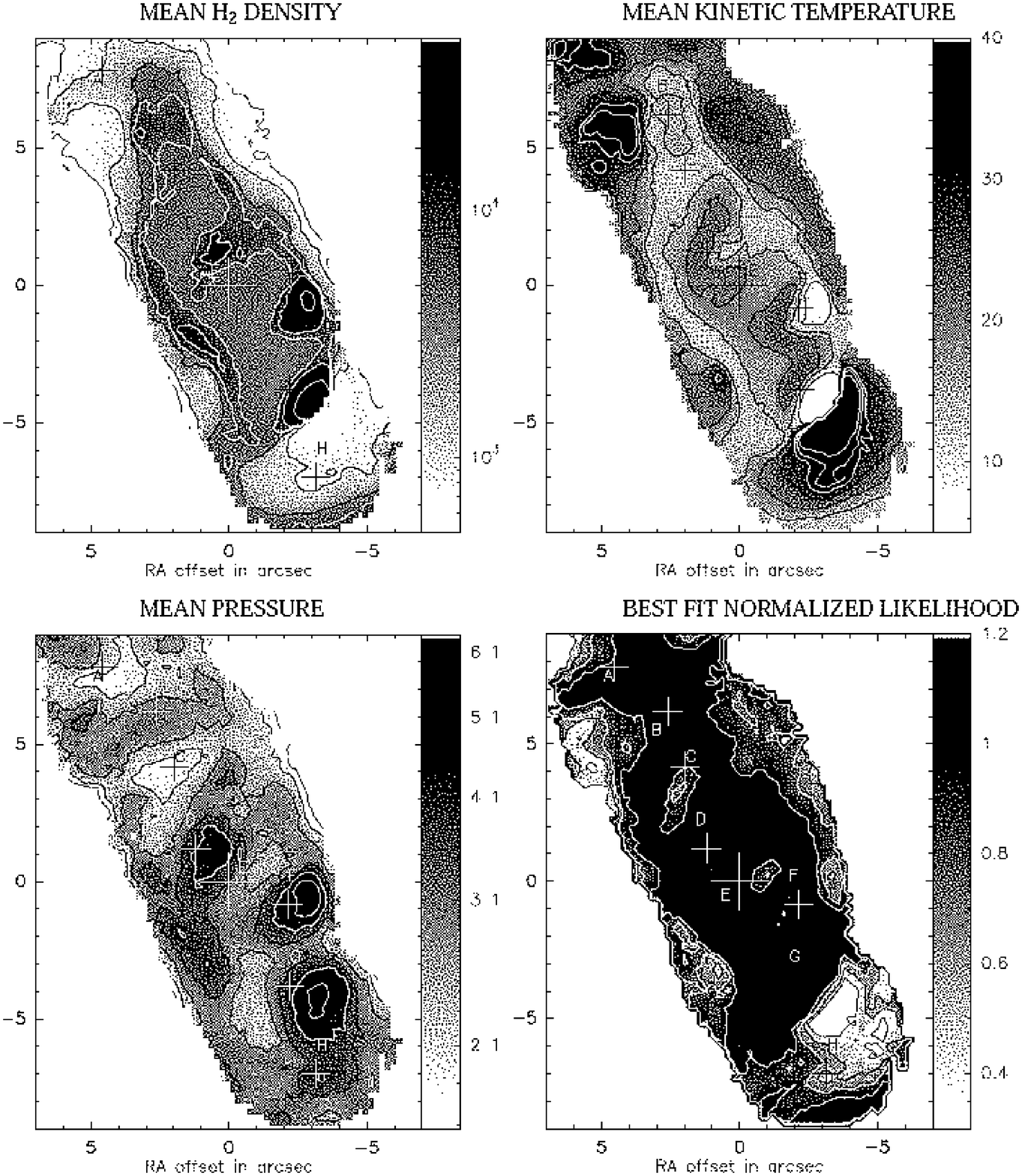}}

\caption{LVG result maps for NGC\,4569. {\bf Upper left}: H$_2$ number density averaged over the 90\% confidence interval of each LVG model. The black contours are at $5\times 10^2$, $10^3$, $3\times 10^3$\,cm$^{-3}$ and the white contours at $5.5\times10^3$, $10^4$,  and $5\times 10^4$\,cm$^{-3}$. {\bf Upper right}: kinetic temperature  averaged over the 90\% confidence interval of each model. The black contours are at 5, 10, 15, 20, 25\,K and the white contours at 30, 35, 40\,K. {\bf Lower left}: pressure averaged over the 90\% confidence interval of each model. The black contours are at $10^{4.1}$, $10^{4.4}$, and $10^{4.7}$\,K\,cm$^{-3}$ and the white contours at $10^{5}$, $10^{5.3}$ and $10^{5.6}$\,K\,cm$^{-3}$.  {\bf Lower right}: the normalized likelihood (or merit), $P/P_{\rm max}$, of the best fit with black contours at 0.35 and 0.55 and white contours at 0.75 and 0.95.}
\label{fig:n4569lvgmaps}
\end{figure*}

\begin{figure*}
\centering

\resizebox{15cm}{!}{\includegraphics{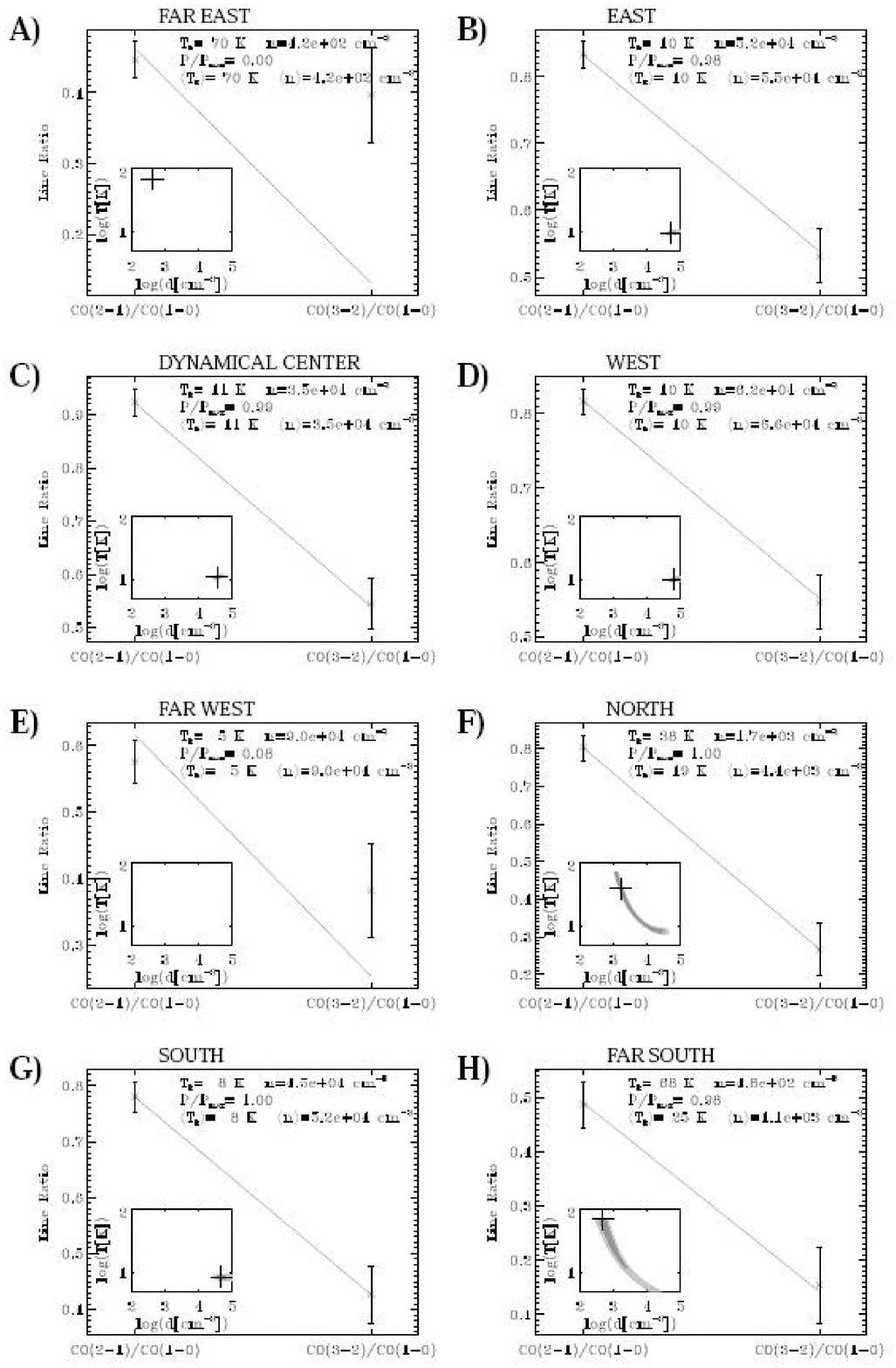}}

\caption{LVG best fits to the line ratios measured in the central region of NGC\,4826 at the eight positions marked in  Fig.\,\ref{fig:n4826mean}. See caption of Fig.\,\ref{fig:N4569lvgfits} for more details.}\label{fig:N4826lvgfits}

\end{figure*}

\begin{figure*}
\centering
\resizebox{17cm}{!}{\includegraphics{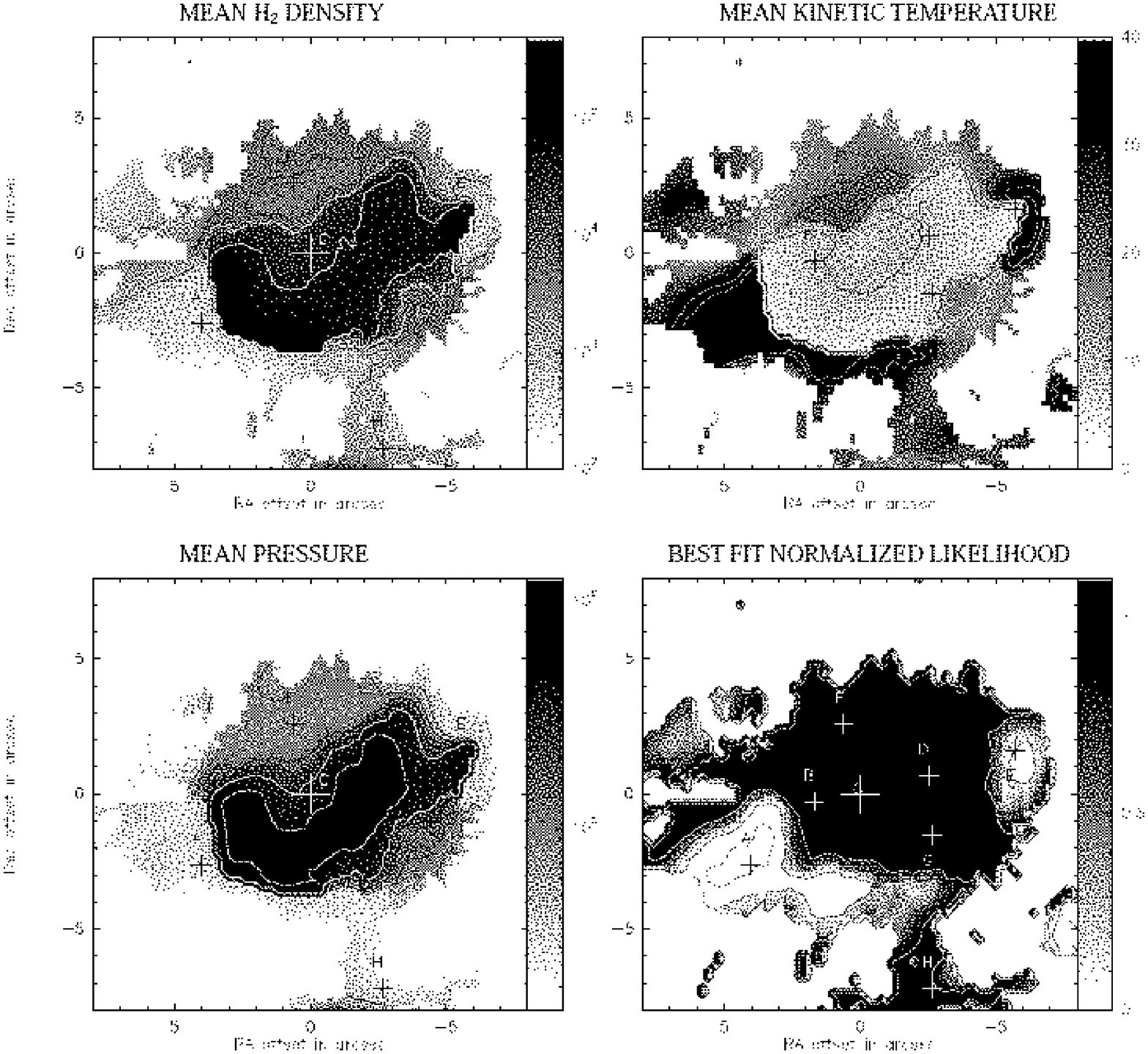}}

\caption{LVG result maps for NGC\,4826. {\bf Upper left}: H$_2$ number density averaged over the 90\% confidence interval of each LVG model. The black contours are at $10^3$ and $5\times 10^3$\,cm$^{-3}$ and the white contours at $10^4$ and  $5\times 10^4$\,cm$^{-3}$. {\bf Upper right}: kinetic temperature  averaged over the 90\% confidence interval of each model. The black contours are at 5, 10, 20\,K and the white contours at 40 and 60\,K. {\bf Lower left}: pressure averaged over the 90\% confidence interval of each model. The black contours are at $10^{4.7}$, $10^{4.9}$, $10^{5.1}$ and $10^{5.3}$\,K\,cm$^{-3}$ and the white contours at $10^{5.5}$ and $10^{5.7}$\,K\,cm$^{-3}$.  {\bf Lower right}: goodness-of-fit with the same contours as in Fig.\,\ref{fig:n4569lvgmaps}.}
\label{fig:n4826lvgmaps}
\end{figure*}

\subsection{LVG model}\label{sec:lvg}

\subsubsection{The procedure}

To interpret the observations of the three CO lines in terms of molecular gas physical conditions we have modeled the line ratios based on the large velocity gradient (LVG) assumption \citep{1973ApJ...180...31S, 1974ApJ...189..441G} using the LVG code in the MIRIAD package. { As opposed to the LTE assumption, the LVG assumption allows us to deal with transitions that are not necessarily thermalized. On the other hand, with the observations available we can only consider a single component.} Although the single component assumption is at odds with the known complexity of the real ISM (even at the 40\,pc linear resolution reached in NGC\,4826), it was successfully used in many extragalactic studies to assess the physical properties of the molecular gas.  For a single component the free parameters are the molecular hydrogen number density, $n$(H$_2$), the kinetic temperature, $T_{\rm k}$, and the ratio of  the CO molecule fractional abundance with the velocity gradient, $X_{\rm CO}/(dV/dR)$. With only two observables (the line ratios) one of the three parameters needs to be fixed. 

We have implemented a maximum likelihood fitting procedure to fit two parameters assuming normal probability distributions for the line ratios. The normalized likelihood of the best fit is used as a goodness-of-fit metric, denoted $\cal M$. It is equal to 1 when the observed line ratios are exactly reproduced by the best fit model and it tends to 0 with increasing discrepancy.

The value of  $X_{\rm CO}/(dV/dR)$ was constrained as follows. We have generated three grids of models for $X_{\rm CO}/(dV/dR)$=$10^{-6}$, $10^{-7}$ and $10^{-8}$\,pc\,(km\,s$^{-1}$)$^{-1}$. Each grid is 
50$\times$50 in size corresponding to 50 values of $T_{\rm k}$ in the range 5 to 200\,K and 50 values of $n$(H$_2$) in the range $10^2$ to $10^5$\,cm$^{-3}$. The intervals in $T_{\rm k}$ and $n$(H$_2$) are logarithmic. The grids are displayed in Fig.\,\ref{fig:lvggrid}.  
To estimate the most appropriate value of $X_{\rm CO}/(dV/dR)$ we computed $\cal M$ for each point of a $R_{32}$-vs-$R_{21}$ grid assuming an uncertainty of 0.02 for $R_{21}$ and 0.04 for $R_{32}$. The distribution of $\cal M$ over the $R_{32}$-vs-$R_{21}$ grid thus obtained was then compared to the $R_{32}$-vs-$R_{21}$ diagram of the galaxies. The three $\cal M$ grids are plotted in Fig.\,\ref{fig:gridgof}. For both galaxies the $X_{\rm CO}/(dV/dR)$ value maximizing the goodness-of-fit over the map is $10^{-7}$\,pc\,(km\,s$^{-1}$)$^{-1}$ similar to that found by \citet{2007ApJ...659..283I} for NGC\,6240 following a similar method; we therefore adopted this value for all the models. 

As can be seen in Fig.\,\ref{fig:lvggrid}, the two line ratios follow a similar distribution over the density-temperature grid, i.e. the isocontours of both line ratios are close to parallel over most of the grid. Hence, in many cases, when the uncertainties are taken into account the solution is degenerate. The confidence intervals typically have arc-shapes similar to the isocontours (see Fig.\,\ref{fig:N4569lvgfits} and \ref{fig:N4826lvgfits}). 
However, a confidence interval does not cover the complete range of densities and temperatures, and the average values over this interval can be used to characterize qualitatively the trend of the model, i.e. whether it tends to favour high or low densities and temperatures.
Thus, instead of considering the best fit solutions, we consider the average values of the temperature and density over all possible solutions, i.e. the average values over the 90\% confidence interval. As the 90\% confidence interval can be unbounded toward high temperatures and/or high densities we fix the maximum possible temperature to 70\,K and the maximum possible density to $10^5$\,cm$^{-3}$. Fixing these upper limits affects the averages but not their qualitative variations with the ratio values. 


\subsubsection{NGC\,4569}

The LVG fits to the eight individual positions marked on the map of Fig.\,\ref{fig:n4569mean} are displayed in Fig.\,\ref{fig:N4569lvgfits}. This figure shows that good LVG solutions exist for all positions but that the uncertainties are large (see the confidence intervals in the insets). This figure also shows how using the average values instead of the best fit gives more stable results. For example, the ratio values at the northern and southern peaks (B and G) are similar and should therefore give similar results. The best fit solutions are very different, the average values are more stable and therefore better adapted to a qualitative description. 


Density and temperature maps obtained by applying the fitting procedure to all pixels are presented in Fig.\,\ref{fig:n4569lvgmaps}. The pressure, obtained by averaging the product, $T\times n$ over the confidence interval, is displayed at the bottom left and the goodness-of-fit map is shown at the bottom right. Gas with density above $5.5\times 10^3$\,cm$^{-3}$ (the lowest white contour) and temperature below 15\,K is distributed around the center in an elongated ring extending from the CO(1--0) northern peak (position B) to the CO(1--0) southern peak (position G). This is consistent with the kinematic model presented in \citetalias{2007A&A...471..113B} where the molecular gas was distributed in an elongated ring  extending from the northern peak to the southern peak and identified as the ILR. Within this ring there are two clumps of high density cold gas, one at the southern CO(1--0) peak (position G) and the other 3$''$ to the west of the nucleus (position F).  

In the central region delineated by the ring there is also a density peak $\sim$1.5$''$ north of the center (near position D) where the temperature is higher than in the ring. On average, in the central region the density is higher than outside the ring. The temperature is higher in the central region than in the ring itself but is not higher than in the region outside the ring. This reversal is not expected, as the central starburst should heat the gas, and this may reflect the limit of the method used here, i.e. averaging the physical parameters over the confidence interval can only give a trend not the absolute values.
The warmest regions are to the east of the northern peak and to the south-west of the southern peak. They may correspond to diffuse hot gas near the points where the orbit crowding causes shocks \citepalias[see orbits in][]{2007A&A...471..113B}. We note however that south of the southern peak the goodness-of-fit is lower (${\cal M} < 0.4$).
There are three pressure peaks coinciding with the three density peaks. 

\subsubsection{NGC\,4826}

The LVG fits to the eight positions marked in the map of Fig.\,\ref{fig:n4826mean} are displayed in Fig.\,\ref{fig:N4826lvgfits}. This figure shows that good LVG solutions exist for all positions except the easternmost and westernmost ones (positions A and E) where $R_{32}$ is too high with respect to  $R_{21}$.

Density and temperature maps obtained by applying the fitting procedure to all pixels are presented in Fig.\,\ref{fig:n4826lvgmaps}. The pressure is displayed at the bottom left, and the goodness-of-fit map is shown at the bottom right.
 The densest (with density $>$$10^4$\,cm$^{-3}$) and coldest ($T\sim 5$\,K) gas is distributed in a semicircle $\sim$60--80\,pc in radius whose center coincides with the dynamical center. Inside the region delineated by the arc the temperature increases from the arc to the north of the dynamical center.
  
 However, as can be seen from the goodness-of-fit map in the lower right panel of Fig.\,\ref{fig:n4826lvgmaps} and as was already noted from the fit to the individual positions A and E (Fig.\,\ref{fig:N4826lvgfits}), the LVG solutions outside the arc along the major axis, i.e. the solutions giving the most diffuse and warm conditions, do not fit the observed ratios well.
 
\section{Discussion}\label{sec:discussion}

\begin{figure}
\centering
\resizebox{7cm}{!}{
\includegraphics[width=7cm, clip=true, viewport=1.9cm 1.7cm 18cm 18cm]{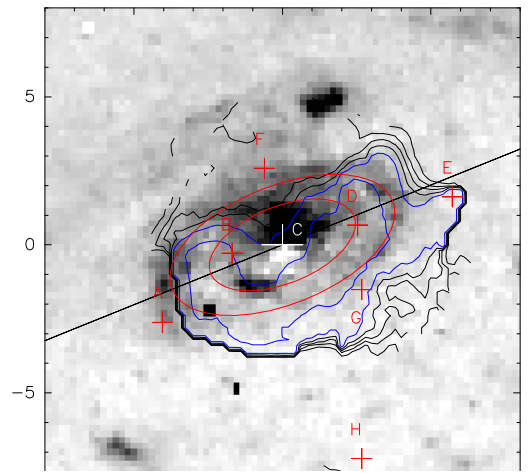}
}
\caption{Pressure contours of NGC\,4826 (same contours as in Fig.\,\ref{fig:n4826lvgmaps} lower left panel) overlaid on the Pa$\alpha$ map. The ellipse corresponds to a ring of 90\,pc radius inclined by 60\,deg. The straight line indicates the major axis of the galaxy.}
\label{fig:n4826arc}
\end{figure}

\begin{figure*}
\centering
\resizebox{16cm}{!}{
\includegraphics{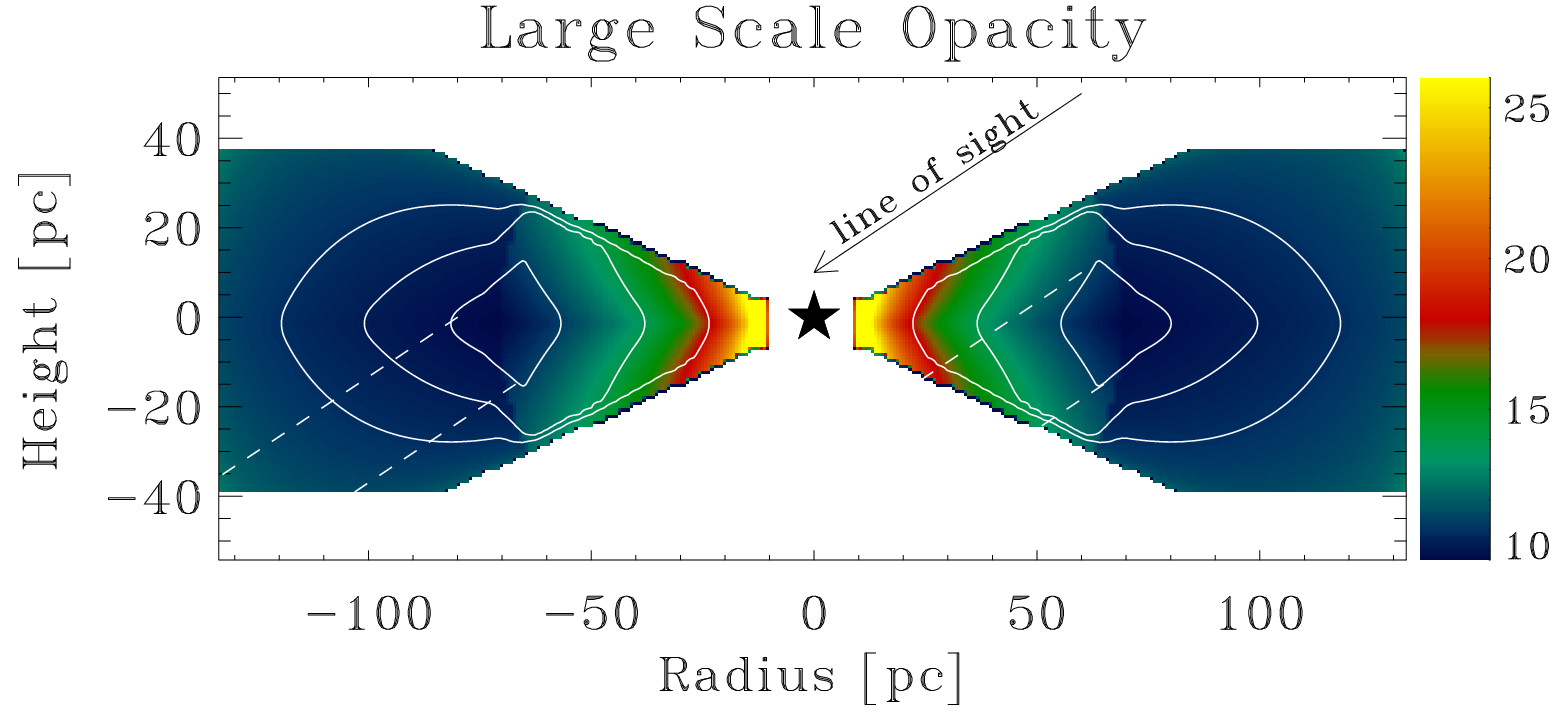}
}
\resizebox{16cm}{!}{
\includegraphics{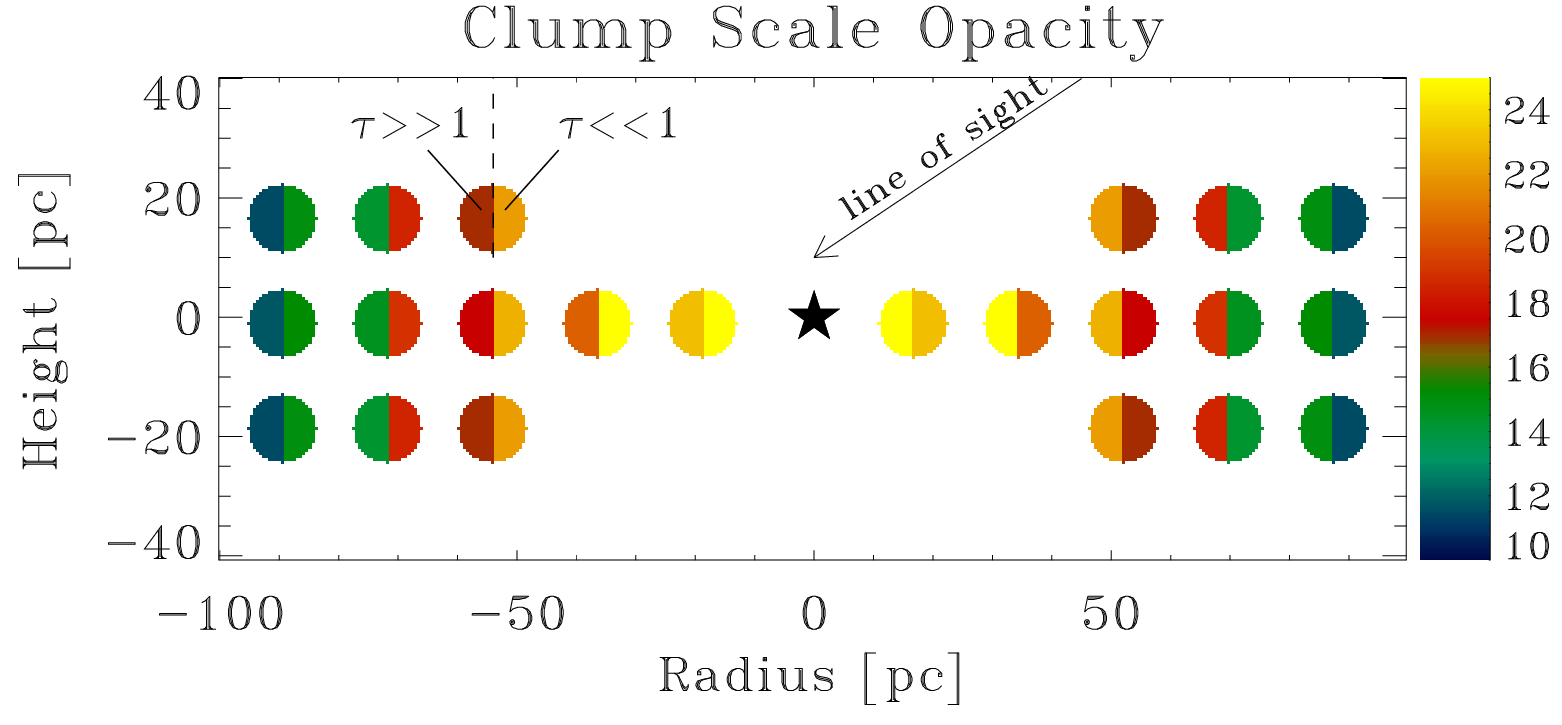}
}

\caption{Sketches illustrating the two opacity effects considered in Section\,\ref{sec:opacity}. In both panels the color scale is related to the temperature in K as shown in the colorbars and the star symbol represents the AGN. The observer is assumed to be looking from the top right, so that the right-hand side of the panels corresponds to the near side and the left-hand side to the far side of the galaxy. The top panel illustrates how the densest and coldest gas distributed in a ring at 70\,pc from the center could partly obscure emission from the dense torus on the near side  and from the galactic disk on the far side. The white contours correspond to density levels. The dashed white lines delineate the regions that would be obscured by the densest gas.   The lower panel illustrates how clumps heated by the nucleus could have inhomogeneous properties with the side exposed to the nucleus being warmer, less dense and less opaque to CO emission than the other side. 
}
\label{fig:torusmodel}
\end{figure*}

\begin{figure*}
\centering
\resizebox{16cm}{!}{\includegraphics{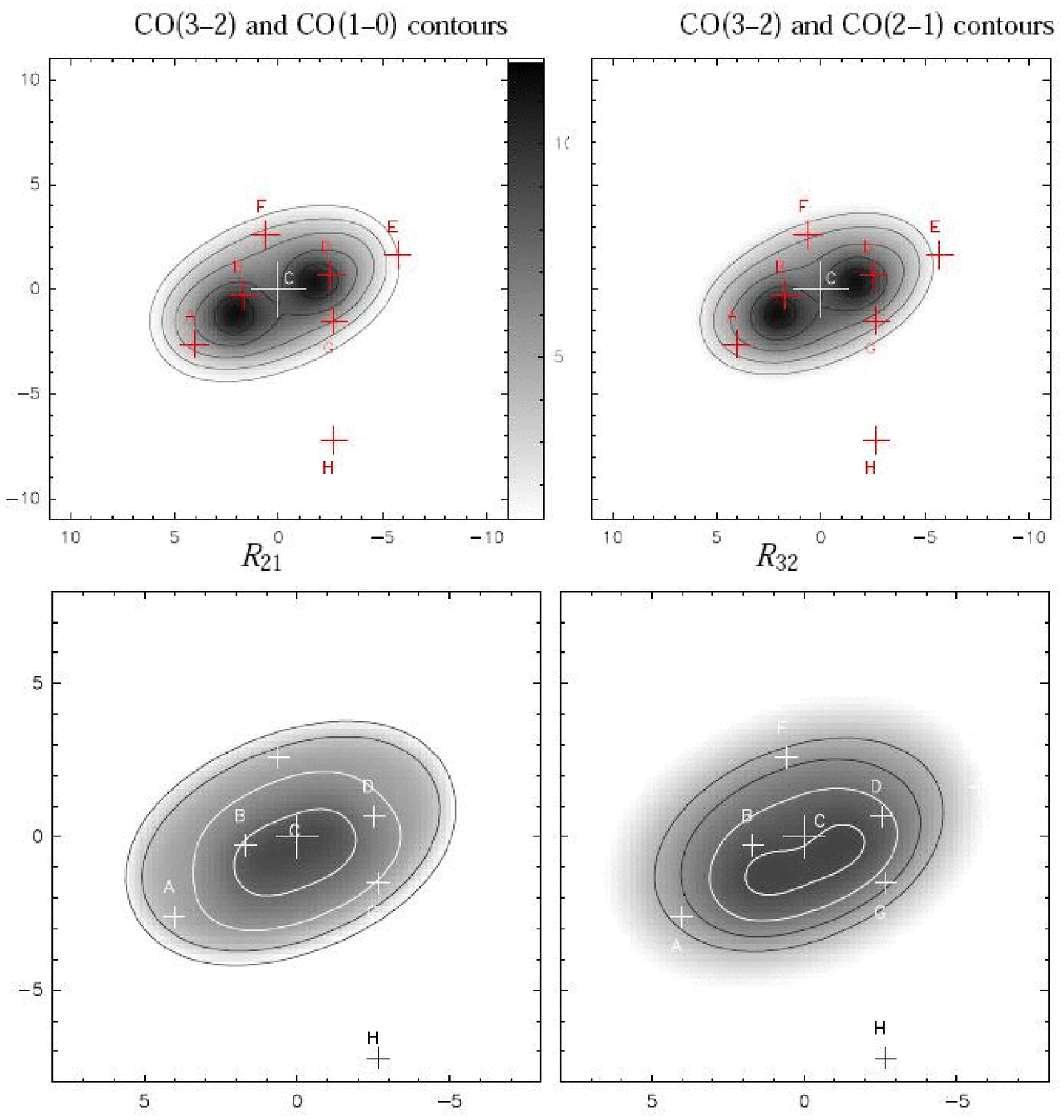}}

\caption{The upper panels show the CO integrated maps produced by the flared disk toy model described in Appendix\,\ref{ap:1}. The CO(3--2) emission is shown in greyscale, and the CO(1--0) (left panel)  and CO(2--1) emission (right panel) are contours. The lower left and right panels show the resulting $R_{21}$ and $R_{32}$ maps, respectively. }
\label{fig:torsim}
\end{figure*}

\begin{figure*}
\centering
\resizebox{16cm}{!}{\includegraphics{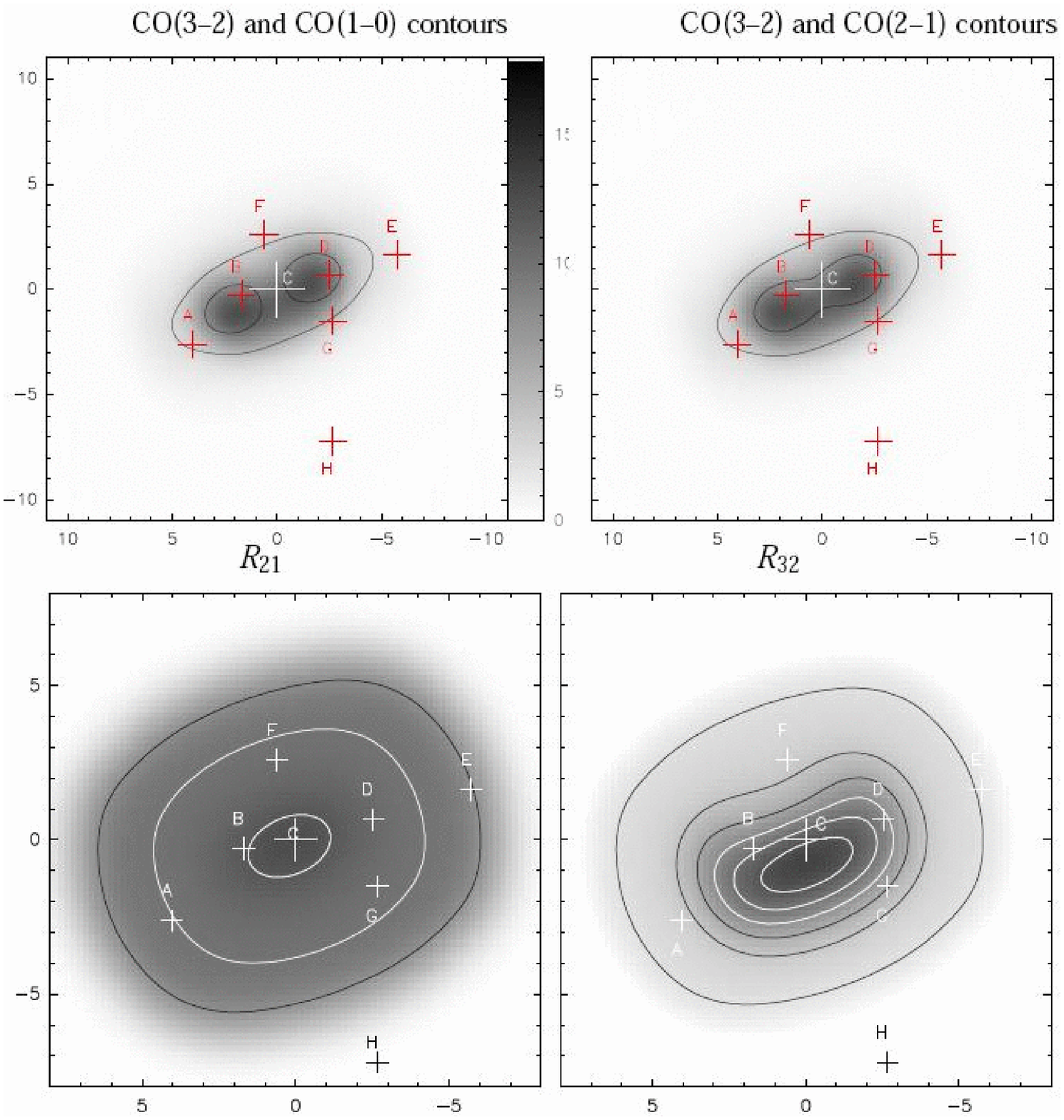}}

\caption{The upper panels show the CO integrated maps produced by the clumpy flared disk toy model described in Appendix\,\ref{ap:2}. The CO(3--2) emission is shown in greyscale, and the CO(1--0) (left panel)    and CO(2--1) emission (right panel) are contours. The lower left and right panels show the resulting $R_{21}$ and $R_{32}$ maps, respectively. }
\label{fig:clumpssim}
\end{figure*}

\subsection{Comparison of the two galaxies}

NGC\,4569 and NGC\,4826 both harbor transition type nuclei. It is therefore interesting to compare the ISM close to the nucleus and look for similarities. 
At first sight, the molecular gas distributions and properties appear to be very different in the two galaxies. In NGC\,4569, the gas is concentrated in an ILR ring and a clump close to the center, and the CO is not very excited. In NGC\,4826, the gas is concentrated in a CND and the gas is more excited. In addition, the strong correlation between the two line ratios near the center in NGC\,4826 is not seen in NGC\,4569. 

However,  the low excitation gas in NGC\,4826 (with $R_{32}$$<$0.3) follows a ratio distribution  similar to that of NGC\,4569 (see the $R_{32}$-vs-$R_{21}$ diagrams Fig.\,\ref{fig:ratioratiolte}), and the difference in linear resolution must be taken into account. It is  4 times higher for NGC\,4826 (40\,pc) than for NGC\,4569 (160\,pc). If NGC\,4826 were observed with the same linear resolution as NGC\,4569 the CND would not be resolved and the ratio distribution in the upper right of the $R_{32}$-vs-$R_{21}$ diagram would be averaged into a small clump in the diagram. Conversely, observing NGC\,4569 with the same linear resolution as NGC\,4826 could well reveal more excited gas. 

Taking into account this observational bias, the most robust similarity between the two galaxies is the relatively low excitation of the CO compared to other galaxy centers.  At the hundred parsec scale, $R_{21}$ and $R_{32}$ can be $>$0.8 \citep{1990ApJ...348..434E, 1991ApJ...372...67E, 1992A&A...264..433B, 1992A&A...265..447W} and $>$1 \citep{2001A&A...373..853D}, respectively, in galaxy centers with enhanced star formation,  and  the ratios lie typically in the range 0.7--1.1 and 0.5--1  around luminous AGNs \citep{1999ApJ...516..114P,2008A&A...491..483P, 2009A&A...493..525I}. At the same scale $R_{21}$ even reaches a value of $\sim$2 in the  prototypical Seyfert galaxy NGC\,1068 \citep{2000ApJ...533..850S} as well as in the Seyfert 1 galaxy NGC\,1097 \citep{2008ApJ...683...70H}.
$R_{32}$ also reaches a value of 2 at the hundred parsec scale in the nucleus of M\,51 \citep{2004ApJ...616L..55M} . 
Finally, $R_{32}$ is also lower than in the inner 200\,pc of the Milky Way (a.k.a.\ the circumnuclear zone, CNZ), where $R_{32}$ has an average value of $\sim$0.8 \citep[see Fig.\,3 of ][]
{2007PASJ...59...15O}. 
Thus, at the scale of previous single dish studies and at the scale of the CNZ, the molecular gas in the centers of  NGC\,4569 and  NGC\,4826 appears to be in a rather low excitation state.
 $R_{21}$ and $R_{32}$ are also lower than the model predictions of \citet{2005ApJ...619...93W} for AGN tori.
 
 
\subsection{Can opacity effects explain the asymmetries in NGC\,4826?}\label{sec:opacity}


 The LVG modeling indicates the presence of a semicircular arc of dense gas close to the nucleus of NGC\,4826. 
 Its center of curvature agrees with the dynamical center and it lies at the south-west of the nucleus.
 In fact, this dense arc seems to reflect asymmetries in the CO and $R_{32}$  maps (Figs.\,\ref{fig:n4826mean} and \ref{fig:n4826ratiomaps}). Indeed, close to the nucleus the emission of the three CO lines is slightly more prominent south of the nucleus and, as was noted in Section\,\ref{sec:lineratios_n4826}, the $R_{32}$ distribution is clearly more prominent to the south.
 
 Interestingly, as can be seen in Fig.\,\ref{fig:n4826arc}, the dense arc seems to coincide with regions of the CND where the Pa$\alpha$ emission is weaker. Another interesting point is that the galaxy major axis seems to be a privileged axis for this anticorrelation, i.e., the Pa$\alpha$ emission is more prominent on the near side (to the north east) of the galactic disk, whereas the semicircular dense arc is on the far side; the minor axis seems to be a symmetry axis for both the Pa$\alpha$ emission and the dense gas distribution. As noted in \citetalias{2003A&A...407..485G}, the dust lane extinction is usually stronger on the near side (an effect that turns out to be spectacular in this galaxy outside the CND), and the Pa$\alpha$ near side/far side asymmetry is opposite to the  one expected for star formation in a density wave. 
 A possible explanation for this Pa$\alpha$ asymmetry is that extinction is not confined to dust lanes in a density wave but rather comes from extraplanar dust around the nucleus \citep[see e.g.][]{2006ApJ...640..612M, 2006ApJ...640..625R, 2008MNRAS.384.1469Q}. In addition, part of the Pa$\alpha$ emission close to the nucleus may come from material above or at the surface of molecular clouds that has been ionized by the nuclear activity rather than by young stars distributed in a spiral.

  The near side/far side asymmetries and the Pa$\alpha$/dense gas anticorrelation suggest that the CO emission as well could be affected by opacity \citep[for a similar conclusion based on intensity ratios averaged over a larger
beam, see also ][]{2009A&A...493..525I}.
 Opacity effects are generally not considered for extragalactic CO observations. Although the low-$J$ CO lines are mainly optically thick, the clumpiness of the ISM and the very low probability to have two clumps on the same line of sight with the same line of sight velocity imply that molecular clouds do not shadow each other, and therefore that CO lines are good column density probes \citep[see e.g. ][]{1986ApJ...309..326D}.
 CO morphologies observed in galaxies are thus assumed to reflect the actual gas distribution.  We see however two possible reasons for opacity to play a role in the CO morphologies observed close to galactic centers. First, the column density is larger and the probability of having several clumps on the same line of sight with the same line of sight velocity may not be negligible anymore. Second, the activity of the nucleus is expected to affect the surrounding ISM and as a result the individual clumps may have strongly inhomogeneous properties fully determined by their distance and orientation relative to the nucleus; the clumps on the far side of the galaxy could thus show very different properties from those of the near side.   Both effects, which we refer to as the large scale opacity effect and the clump scale opacity effect, respectively, are sketched in Fig.\,\ref{fig:torusmodel}. We investigate them in more detail in the following.
 
\subsubsection{Large scale opacity effect}
  
 
 The molecular gas disks in active nuclei are thought to flare out to radii of order 10\,pc,
 forming the putative obscuring torus required by the unified AGN model \citep[see e.g. ][and references therein]{2008NewAR..52..274E}.  How this structure connects to the outer disk and whether it is embedded in a larger toroidal structure is still uncertain. The hydrodynamical simulations by \citet{2005ApJ...619...93W} suggest that such a large molecular structure could extend out to several tens of parsecs and this seems to be confirmed by observations of the near-IR molecular hydrogen lines in nearby AGNs by \citet{2009ApJ...696..448H}. 
 
 If the densest and coldest gas is distributed in a ring inside such a large toroidal structure and is able to partly absorb CO emission along the line of sight, then the disk beyond the ring would be partly absorbed on the far side and the warmer inner regions of the torus would be partly absorbed on the near side (see sketch in the top panel of Fig.\,\ref{fig:torusmodel}). In other words, the line of sight toward the torus on the near side would be more contaminated by emission from the disk and would not be seeing all the emission from the inner regions of the torus. The torus would be more clearly visible on the far side.

 
 
In order to check whether this effect could be 
 compatible with the observations we consider a toy model of a flared disk (see details in Appendix\,\ref{ap:1}).
 The resulting integrated CO maps and line ratio maps are presented in Fig.\,\ref{fig:torsim}. 
Although this simple model cannot reproduce exactly the integrated maps shown in Fig.\,\ref{fig:n4826mean}, it can partly reproduce the morphology inside the CND. In particular, the emission does not peak at the nucleus but to the east and west of the nucleus, and there is more emission to the south (the far side) than to the north of the nucleus. Also, the general morphology does not change significantly in the three CO lines. The main differences between the $R_{21}$  and  $R_{32}$ maps noted in Section\,\ref{sec:lineratios_n4826} are also reproduced by the model. The $R_{32}$ distribution is more extended along the major axis and is more prominent to the south of the nucleus (on the far side) than the  $R_{21}$ distribution.


 \subsubsection{Clump scale opacity effect}
 
Alternatively, the asymmetry could result from the inhomogeneity of the clumps forming the toroidal structure. The side of the clumps exposed to the nucleus would be warmer and more diffuse than the other side. The dense side would be more opaque and it would hide the warmer side to the observer depending on the angle between the line of sight and the line subtended by the clump and the nucleus (see sketch in the lower panel of Fig.\,\ref{fig:torusmodel}). A similar "external heating" scenario was already discussed by \citet{1998IAUS..184..221B} and \citet{2000PhDT.........6B} and by \citet{2001ApJ...551..687M} to interpret asymmetries of molecular line emission in NGC\,1068 and  IC\,342, respectively.
A similar argument was also invoked and formalized  by \citet[][]{2008ApJ...685..160N, 2008ApJ...685..147N} to interpret the infrared observations of AGNs.  To test whether such an effect could be consistent with the observations we have distributed clumps in the same toroidal structure as the one used in the model of the previous subsection. 
The resulting CO maps as well as the ratio maps are presented in Fig.\,\ref{fig:clumpssim}. They show that this simple model can  also reproduce the slight CO emission excess on the far side and the differences between the two ratio maps noted in Section\,\ref{sec:lineratios_n4826}.
 
\subsubsection{A common property of active galaxies?}

We note that in NGC\,4569 as well, the CO emission and the $R_{32}$ distributions close to the nucleus are more prominent on the far side (east side). In fact, although this remained unnoticed for most of them, several nearby active galaxies show a slight near side/far side asymmetry close to their nuclei with CO emission excess on the far side as well; this is for example the case for Centaurus A \citep{2009ApJ...695..116E}, NGC\,4945 \citep{2007ApJ...670..116C}, NGC\,3227 \citep{2000ApJ...533..826S}, and NGC\,1068 \citep{2000ApJ...533..850S}. A detailed statistical study of CO asymmetries in nearby AGNs is beyond the scope of this paper, but these few cases suggest that near side/far side asymmetries could be common in AGNs.

While both torus models described above can reproduce the asymmetry observed, the first model requires a high inclination, a very specific geometry, and a high volume filling factor of the dense gas. The second model is more robust to various inclinations (although it also decreases with decreasing inclination) and configurations; it would therefore provide the best explanation for a putative generalized near side/far side asymmetry in AGNs. An alternative opacity effect was proposed by \citet{2003A&A...412..615G}  to explain the asymmetry in NGC\,1068. They invoked CO line absorption by a thick disk, and in their model the asymmetry resulted from a tilt between the molecular disk and the Compton thick absorber. This model however does not explain why the asymmetry should coincide with a far side/near side asymmetry.

\section{Conclusion}\label{sec:conclusion}

We have observed the emission from the first three $^{12}$CO $J$-lines in the nuclei of two nearby galaxies, NGC\,4569 and NGC\,4826, with the Plateau de Bure interferometer  and the SubMillimeter Array. Combining these data allows us to compare the emission in the three lines and to map the line ratios $R_{21}$ and $R_{32}$ at a resolution of $\sim$2$''$, i.e. a linear resolution of 160\,pc for NGC\,4569 and 40\,pc for NGC\,4826. We interpret the line ratios through radiative transfer modeling and map the physical conditions of the molecular gas. The main results of this study can be summarized as follows:
\begin{itemize}
\item In both galaxies the three lines are similarly distributed over the data cubes, i.e. both spatially and in velocity.
\item In both galaxies the line ratios seem to be lower than in other galaxy centers.
\item In both galaxies the gas with $R_{32}$ greater than 0.2 is confined to Pa$\alpha$ emitting regions detected by NICMOS. $R_{21}$ instead reaches high values away from the   Pa$\alpha$ emitting regions detected by NICMOS. 
\item According to a pseudo-LTE model, the molecular gas in NGC\,4569 is cold, mainly optically thick in the CO(1--0) and (2--1)  lines and with a fraction less than 50\% optically thin in the CO(3--2) line. LVG modeling suggests the presence of an elongated ring of cold and dense gas coinciding with the ILR. There are two denser and colder clumps along this ring and one denser and warmer clump close ($\sim$160\,pc) to the nucleus.
\item More excited gas is resolved in the circumnuclear disk (CND) of NGC\,4826. According to the pseudo-LTE model this corresponds to warmer gas with a $\sim$50$\%$ of the CO(3--2) emission being optically thin. The LVG modeling indicates the presence of a semicircular arc of dense and cold gas centered on the dynamical center and $\sim$70\,pc in radius. The gas temperature increases and the density decreases toward the center. 
\item NGC\,4826 shows a near side/far side asymmetry in the CO, Pa$\alpha$ and $R_{32}$ distributions.   This asymmetry suggests that the gas opacity could play a role in the CO morphology. It can be reproduced if the cold dense gas is distributed in a ring of 70\,pc radius embedded in a flared toroidal structure and can partly absorb the CO emission, or if the clumps forming the toroidal structure are inhomogeneously heated by the nuclear activity and are partly optically thick. 
\item Several published CO maps of nearby active galaxies seem to show a slight CO emission excess on the far side of the nuclear region similar to the one observed in NGC\,4826 suggesting that this could be a common property of active galaxies. If confirmed, this would favor the interpretation based on the model with inhomogeneous clumps heated by the nuclear activity.
\end{itemize}
This study demonstrates the feasibility of mapping the molecular gas physical properties in the centers of nearby galaxies based on multitransition CO observations with current interferometers. It also shows how important these observations are for correctly interpreting the CO emission and thus understanding the processes at work in nuclear regions. While the results derived in this study are  still mainly qualitative and limited by uncertainties and resolution, they suggest new areas of investigation. Analyzing multitransition interferometric CO observations for more nearby galaxies with various inclinations should provide a statistical basis for investigating the unified AGN paradigm. More importantly, with ALMA, such observations will be greatly improved in terms of sensitivity (about one order of magnitude), resolution (one order of magnitude) and imaging quality (more complete {\it uv} coverage). Furthermore, it will be possible to observe more CO transitions with a single array and thus, to better constrain the models and to derive more quantitative results on the ISM properties close to galaxy centers. For example it will be possible to keep the CO abundance as a free parameter and thus resolve the possible AGN influence on this parameter. It will also be possible to resolve the clumpy structure of the CNDs and investigate in more detail the role of pressure in these structures. 

\begin{acknowledgements}
We are much indebted to Christine Wilson and Frank Israel for very kindly providing us with their single dish CO(3--2) data.
FB thanks the Max-Planck-Institut f\"ur Radioastronomie in Bonn, where he was a postdoc at the time of the observations, for supporting the travel to the SMA.
 We thank the scientific and technical staffs at the SMA for their help at the observatory. The collaboration between LERMA and ASIAA was supported by the ORCHID French-Taiwanese Hubert Curien Program.
\end{acknowledgements}

\bibliographystyle{aa}
\bibliography{nuga-sma}

\Online

\appendix

\section{Large Torus Models}
\begin{figure}
\centering
\resizebox{8cm}{!}{
\includegraphics{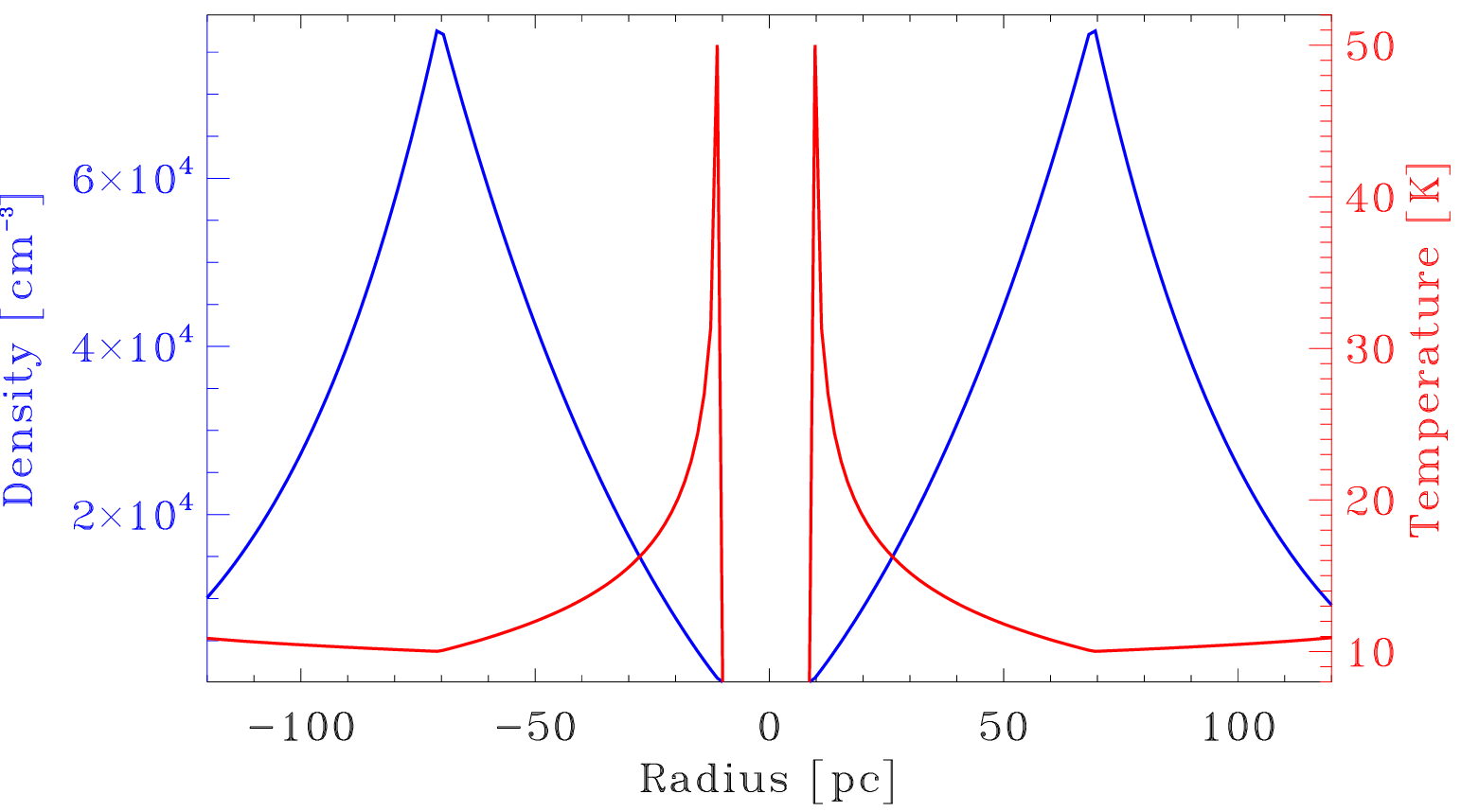}
}
\caption{Radial profiles of the molecular gas density (blue) and temperature (red) used in the models.}
\label{fig:torprofiles}
\end{figure}
This appendix describes in more detail the two models discussed in Section\,\ref{sec:opacity}.  We emphasize that these models are mainly geometrical toy models and they do not rely on any realistic simulations of the ISM. Their physical parameters were not fitted to the observations. These models are only meant to test whether large scale or clump scale  opacity effects in a generic toroidal structure could be consistent with the near side/far side asymmetry observed in NGC\,4826.
\subsection{Model with large scale opacity}\label{ap:1}
In this model the clumps are assumed to be homegeneous and the only parameter related to the clumpiness of the medium is the volume filling factor. The sizes and shapes of the individual clumps are not taken into account. The density and temperature radial profiles used are represented in Fig.\,\ref{fig:torprofiles}.
The density increases exponentially with radius from $n_{\rm min}=10^{2.5}$\,cm$^{-3}$ at 10\,pc to $n_{\rm max}=10^{4.9}$\,cm$^{-3}$ at 70\,pc and it decreases exponentially with height above the galactic plane. Inside 70\,pc the temperature is related to the density by $T \propto n^{\alpha}$ with $T_{\rm max}=50$\,K at 10\,pc and $T_{\rm min}=10$\,K at 70\,pc, i.e. ${\alpha}=-0.3$. Outside 70\,pc the density decreases exponentially with radius down to $n_{\rm out}=10^{2}$\,cm$^{-3}$ and the temperature is kept constant at 12\,K. Overall the density and temperature profiles define a cold dense ring at 70\,pc. The gas volume filling factor is proportional to the gas density (where the gas is denser, it fills more of a given volume). We assume that gas with a density higher than $0.6\times n_{\rm max}$ has a volume filling factor of 20\% and can absorb the CO line emission from gas further away on the line of sight having the same line of sight velocity. We assume circular rotation and a gaussian line profile with a velocity dispersion of 15\,km\,s$^{-1}$ outside 70\,pc and 30\,km\,s$^{-1}$ inside. 
The CO line intensities are computed with the LVG model. The resulting integrated CO maps and line ratio maps are presented in Fig.\,\ref{fig:torsim}. 

\subsection{Model with clump scale opacity}\label{ap:2}

In this model the individual clumps are assumed to be spherical and inhomogeneous and they cannot obscure each other. In each clump the hemisphere exposed to the nucleus is assumed to be warmer and less dense than the other hemisphere. In addition, the warm hemisphere is assumed to be optically thin in the CO(3--2) line, while the other hemisphere is assumed to be optically thick. CO(1--0) and (2--1) lines are assumed to be optically thick everywhere. The density and temperature profiles of the warm hemisphere of the clumps  as  a function of the distance to the nucleus and the height above the plane are the same as in the previous model. The temperature profile of the cold hemisphere follows the same exponential profile with the same lower value but the maximum value is fixed to 30\,K instead of 50\,K. Thus, the difference in temperature between the warm and cold hemispheres increases with the temperature and reaches 20\,K at the maximum. The density of the gas in the clumps is related to the temperature as in the previous model and the number density of clumps is assumed proportional to this gas density.
The line intensities are computed assuming LTE. 
The resulting CO maps as well as the ratio maps are presented in Fig.\,\ref{fig:clumpssim}.

\end{document}